\documentclass[showpacs,amssymb,twocolumn,floatfix,pre,aps]{revtex4-1}

\usepackage{amssymb,amsfonts,bm}
\usepackage{graphics,graphicx}

\begin{document}

\title{Phase diagram and critical properties of Yukawa bilayers}

\author{Igor Trav{\v e}nec}
\affiliation{Institute of Physics, Slovak Academy of Sciences, 
D\'ubravsk\'a cesta 9, 84511 Bratislava, Slovakia}
\author{Ladislav \v{S}amaj}
\affiliation{Institute of Physics, Slovak Academy of Sciences, 
D\'ubravsk\'a cesta 9, 84511 Bratislava, Slovakia}

\begin{abstract}
We study the ground-state Wigner bilayers of pointlike particles 
with Yukawa pairwise interactions, confined to the surface of two parallel 
hard walls at dimensionless distance $\eta$. 
The model involves as limiting cases the unscreened Coulomb potential
and hard spheres. 
The phase diagram of Yukawa particles, studied numerically by 
Messina and L\"owen [Phys. Rev. Lett. 91 (2003) 146101], exhibits five 
different staggered phases as $\eta$ varies from 0 to intermediate values.
We present a lattice summation method using the generalized Misra functions 
which permits us to calculate the energy per particle of the phases with 
a precision much higher than usual in computer simulations.
This allows us to address some tiny details of the phase diagram.
Going from the hexagonal phase I to phase II is shown to occur at $\eta=0$,
which resolves a longtime controversy.
We find a tricritical point where Messina and L\"owen suggested 
a coexistence domain of several phases which was suggested to divide 
the staggered rhombic phase into two separate regions.
Our calculations reveal one continuous region for this rhombic phase
with a very narrow connecting channel. 
Further we show that all second-order phase transitions are of 
mean-field type.
We also derive the asymptotic shape of critical lines close to the Coulomb 
and hard-spheres limits.
In and close to the hard-spheres limit, the dependence of the internal 
parameters of the present phases on $\eta$ is determined exactly.  
\end{abstract}


\date{\today}

\maketitle

\section{Introduction} \label{Sec1}
Most organic or anorganic surfaces of mesoscopic objects (macromolecules 
or colloids) become charged when immersed in polar solvents such as water.
These solvents provide favorable environments for free charges
(``counter-ions'' to the charged surfaces) which intermediate 
an effective interaction among the mesoscopic objects.
At low temperatures, and in particular at $T=0$ when the system is in its
ground state, counter-ions between two charged plates crystallize into 
bilayer Wigner structures which are important in understanding anomalous 
phenomena such as like-charge attraction or overcharging 
\cite{Lau,Grosberg,Levin,Naji,Samaj}. 
Bilayer Wigner crystals describe several real physical systems in
condensed and soft matter, such as semiconductors \cite{Fil}, quantum dots
\cite{Imamura} and dusty plasmas \cite{Teng}. 
Confined systems of charged colloidal particles were reviewed recently 
\cite{review}, both from experimental and theoretical point of view.

\begin{figure}[thb]
\begin{center}
\includegraphics[clip,width=0.35\textwidth]{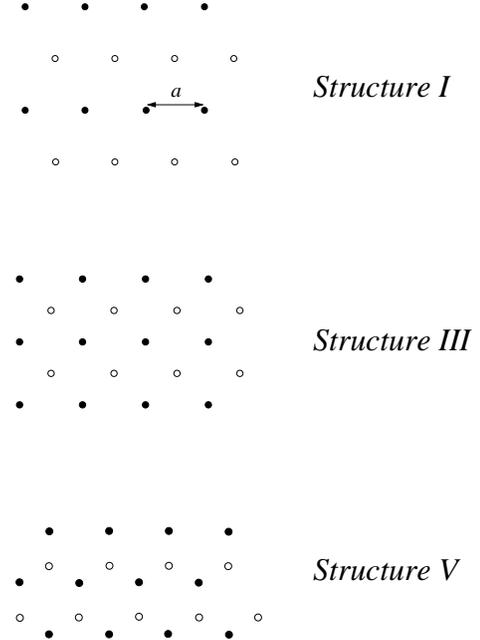}
\caption{Rigid structures I, III and V of particles on two parallel plates; 
open and filled symbols correspond to particle positions on the opposite 
layers.}
\label{fig:Structures}
\end{center}
\end{figure}

\begin{figure}[thb]
\begin{center}
\includegraphics[clip,width=0.40\textwidth]{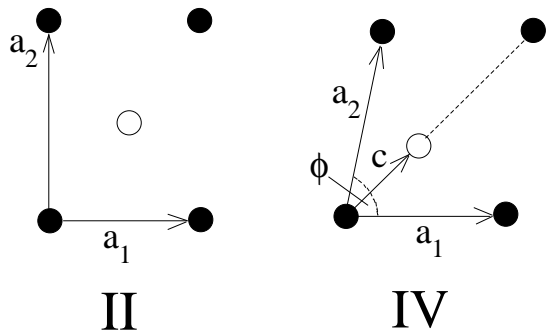}
\caption{Soft structures II and IV.}
\label{fig:Structures24}
\end{center}
\end{figure}

From the particle models studied in this paper, we start with 
the neutral Coulomb system of say elementary pointlike charges $-e$ with 
$1/r$ interaction between two parallel plates of the same homogeneous surface 
charge density $\sigma e$ at distance $d$, the phase diagram at $T=0$ 
depends on a single dimensionless parameter $\eta = d\sqrt{\sigma}$.
According to the Earnshaw theorem \cite{Earnshaw}, particles will stick
symmetrically on the surface of the plates.
Five distinct phases were detected to be stable, i.e. providing global
minimum of the energy, as $\eta$ is changing from 0 to $\infty$ 
\cite{Falko,Esfarjani,Goldoni,Schweigert,Weis}.
The lattice structures are the same on both plates and they are shifted 
laterally with respect to one another.
Structures I, III and V are rigid (Fig. \ref{fig:Structures}), i.e. they
have fixed ($\eta$-independent) primary cells.
Structures II and IV are soft (Fig. \ref{fig:Structures24}), 
the shape of their primary cells is varying with $\eta$.

\bigskip
\bigskip

Structures I, II and III correspond to the staggered rectangular lattice, 
see Fig. \ref{fig:Structures24} left.
The primitive translation vectors of Bravais lattice are 
\begin{equation} \label{structII}
{\bm a}_1 = a(1,0) , \quad {\bm a}_2 = a(0,\Delta) , \quad
a = \frac{1}{\sqrt{\sigma \Delta}} . 
\end{equation}
The lattice spacing $a$ within one layer is determined by 
the electroneutrality condition.
Two rectangular structures, one within each layer, are shifted 
with respect to each other by the vector 
\begin{equation} \label{vectorc}
{\bm c} = \alpha ({\bm a_1}+{\bm a_2}) 
\end{equation}
with $\alpha=1/2$.
The rigid structure I has the aspect ratio $\Delta=\sqrt{3}$ and it arises 
for $\eta = 0$ because the two layers merge into a Wigner monolayer which 
is known to be hexagonal (or, equivalently, equilateral triangular) 
\cite{Bonsall77}.
Structure III consists of a square lattice with $\Delta=1$.
Phase II with $\sqrt{3}>\Delta>1$ interpolates continuously between 
structures I and III. 

Phase IV consists of two staggered rhombus lattices 
(Fig. \ref{fig:Structures24} right). 
One rhombus structure has the angle $\phi$ between the primitive translation 
vectors
\begin{equation}
{\bm a}_1 = a(1,0) , \quad {\bm a}_2 = a(\cos\phi,\sin\phi) , \quad
a = \frac{1}{\sqrt{\sigma \sin\phi}} . 
\end{equation}
In general, this phase has two variants according to the lateral shift 
(\ref{vectorc}) between the opposite sublattices. 
The version IVA has $\alpha=1/2$ whereas for IVB the shift parameter 
$1/3<\alpha<1/2$.
For Coulomb bilayers, only phase IVA takes place.

Phase V corresponds to two shifted hexagonal lattices.
In a single layer, the elementary cell is the rhombus with the
primitive translation vectors  
\begin{equation}
{\bm a}_1 = a(1,0) , \quad {\bm a}_2 = \frac{a}{2} (1,\sqrt{3}) , \quad
a = \frac{\sqrt{2}}{3^{1/4}\sqrt{\sigma}} . 
\end{equation} 
The lateral shift between the opposite lattices ${\bm c}$ is given by
(\ref{vectorc}) with $\alpha=1/3$.

The transitions between phases ${\rm II}\to {\rm III}$ and 
${\rm III}\to {\rm IV}$ are continuous (of second order), 
while the transition ${\rm IV}\to {\rm V}$ is discontinuous (of first order).
In order to describe these phase transitions, a new analytic approach
to Coulomb bilayers was proposed in Ref. \cite{Samaj1}.
Using a series of transformations with Jacobi theta functions, the
energy of the five phases was expressed as series of generalized Misra 
functions which converge very quickly.
Near critical points, the generalized Misra functions can be expanded
easily in powers of the order parameter and the corresponding energies 
posses the Ginsburg-Landau form.
This allows one to specify the critical points with an arbitrary 
prescribed accuracy and to derive the mean-field critical behavior of 
the order parameter.
Also the existence of phase I at $\eta=0$ only was confirmed.
This result was in contradiction with numerical approaches like 
Ewald technique \cite{Goldoni} and Monte Carlo simulations \cite{Weis}
which predicted an extremely small, but finite, stability interval
of $\eta$'s for phase I.
 
Colloidal particles or particles in highly charged dusty plasmas usually 
interact via Yukawa potential \cite{Nunomura} due to the Coulomb potential 
screening by additional microions in the system.
The Yukawa pair potential of particles at distance $r$ is defined by
\begin{equation} \label{yukawa}
V(r)=V_0\ \frac{{\rm e}^{-\kappa r}}{\kappa r} ,
\end{equation}
where $\kappa$ is the inverse screening length and the amplitude
$V_0=Z^2\kappa\exp{(\kappa R)}/\epsilon(1+\kappa R)^2$, with $Z$ being 
the charge of one particle and $\epsilon\approx \epsilon_0$ is 
the permittivity for dusty plasma. 
When $\kappa$ is large, $R$ becomes the radius of a hard sphere 
as $V(r)\propto\exp{[\kappa (R-r)]}$ is exponentially large 
for $r < R$ and negligible otherwise.
The relation $V_0\propto \kappa$ keeps the limit $\kappa\to 0$ of 
Eq. (\ref{yukawa}) finite, yielding the proper Coulomb formula.
Thus the limiting cases $\kappa\to 0$ and $\kappa\to \infty$ correspond to 
the unscreened Coulomb and hard-spheres interaction potentials, respectively.
We shall work in units of $V_0=1$.
For two parallel plates at distance $d$, the phase diagram depends on 
two dimensionless parameters 
\begin{equation}
\eta = \sqrt{\sigma} d , \qquad \lambda=\kappa d . 
\end{equation}

A system of hard-sphere particles between two parallel hard plates was studied 
by computer simulations in the past \cite{Murray,Schmidt,Neser,Fortini};
numerical methods were reviewed recently in Ref. \cite{Mazars11}.
For small values of $\eta$, the ground-state crystal structures involve 
Wigner bilayers I-V, including phase IVB with two varying parameters 
$\phi$ and $\alpha$.
For large values of $\eta$, phase-V bilayer transforms itself to crystalline 
multilayers, with particles entering the region between the plates,
such as multiple square and hexagonal layers \cite{Murray}, 
rhombic \cite{Schmidt} and prism superlattices \cite{Neser}.  

A similar phase diagram was obtained for the general Yukawa potential.
For small values of $\eta$, although Earnshaw theorem \cite{Earnshaw}
does not apply to Yukawa particles, the particles stick symmetrically
to plates and with increasing $\eta$ they constitute successively Wigner I-V 
bilayers \cite{Messina}. 
In the region of large values of $\eta$, in close analogy with confined 
hard spheres, some of the particles will move in the interior of the domain
between the plates and create multilayers \cite{Oguz}.

In this paper, we shall concentrate on Wigner bilayers of pointlike
particles interacting via Yukawa potential.
The original numerical work of Messina and L\"owen \cite{Messina} determined 
the phase diagram of the Yukawa system which exhibits single and double 
reentrant transition.
We shall apply a straightforward extension of the recent analytic method 
\cite{Samaj1} which provides us with high precision calculations 
to shed more light on important tiny details of the phase diagram.
For any $\lambda$, the transition from phase I to II is shown to occur 
directly at $\eta=0$, which solves a longtime controversy.
We recall that this scenario was anticipated only in the hard-spheres limit
$\lambda\to\infty$ \cite{Messina}.
We find a tricritical point where Messina and L\"owen suggested 
a coexistence domain of several phases which should divide one staggered 
rhombic phase into two separate regions.
Our calculations reveal one continuous region for this rhombic phase
with a narrow connecting channel. 
Closed-form formulas for critical lines between various phases,
expressed in terms of generalized Misra functions, permit us to
determine the asymptotic Coulomb $\lambda\to 0$ and hard spheres
$\lambda\to\infty$ shapes of these lines.
The expansions of the structure energies around second-order transition
points and the determination of the order parameter can be done analytically, 
which enables us to derive the critical behavior of the Ginzburg-Landau type. 
In and close to the hard-spheres limit, the $\eta$-dependence of the internal 
parameters of the phases is determined exactly.  
 
The paper is organized as follows. 
In Sec. \ref{Sec2}, we derive the expression for the energy per particle 
of phase II (phases I and III being its special cases) in terms of 
the generalized Misra functions.
The fact that going from phase I to II occurs at $\eta=0$ is shown 
in Sec. \ref{Sec3}.
The second-order transition between phase II and III and the corresponding 
mean-field critical behavior are described in detail in Sec. \ref{Sec4}.
The expression for the energy of phase IVB (with phases IVA and V as 
its special cases) is derived in Sec. \ref{Sec5}.
The second-order transition between phases III and IVA is described 
in Sec. \ref{Sec6}.
The first-order transitions between phases IVA-V, IVA-IVB and IVB-V
are discussed in Sec. \ref{Sec7}. 
The dependence of the energy on the dimensionless distance $\eta$,
for fixed values of $\lambda$, is the subject of Sec. \ref{Sec8}.
The $\eta$-dependence of the internal structure parameters of phases
present in and near the hard-spheres limit is derived in Sec. \ref{Sec9}.  
Sec. \ref{Sec10} is the Conclusion.
Auxiliary formulas for the generalized Misra functions and for the
critical lines are given in Appendices A-D.

\section{Energy of structures I, II and III} \label{Sec2}
We aim at deriving the interaction energy per particle $E_{\rm II}$ for 
the structure II with the aspect ratio $\Delta$, phases I and III being 
its special cases with $\Delta = \sqrt{3}$ and $\Delta=1$, respectively.
The energy consists of two parts: the intralayer energy $E_{\rm intra}$
sums the contributions from all particles in the same layer as
the reference one while the interlayer energy $E_{\rm inter}$ involves
all particles from the opposite layer. 
To express the energy per particle as a quickly convergent series,
we shall apply a three-step method from Ref. \cite{Samaj1}.

The occupied lattice sites within one layer are numbered as
${\bm r} = j{\bm a}_1 + k {\bm a}_2$ where the primitive vectors
${\bm a}_1$ and ${\bm a}_2$ are defined in (\ref{structII}) and
$j, k$ run over all integers, except for the reference site $(0,0)$.
The intralayer interaction of a reference particle is thus given by
\begin{equation}
E_{\rm intra} = \frac{1}{2} \sum_{(j,k)\ne (0,0)}
\frac{\exp\left( -\kappa a\sqrt{j^2+k^2\Delta^2}\right)}{
\kappa a \sqrt{j^2+k^2\Delta^2}} .
\end{equation}
To evaluate lattice sums of Yukawa potentials, we shall often use
the integral representation (see e.g. \cite{Mazars11})
\begin{equation} \label{yukawatrans}
\frac{{\rm e}^{-\kappa r}}{\kappa r} = \frac{1}{\kappa\sqrt{\pi}}
\int_0^{\infty} \frac{{\rm d}t}{\sqrt{t}}
\exp\left( -\frac{\kappa^2}{4t}-r^2 t \right) . 
\end{equation}
The intralayer energy per particle is then expressible as
\begin{eqnarray}
E_{\rm intra} =  \frac{1}{2a\kappa\sqrt{\pi}}\int_0^\infty 
\frac{{\rm d}t}{\sqrt{t}} {\rm e}^{-\frac{\kappa^2 a^2}{4t}}
\left( \sum_{j,k}{\rm e}^{-j^2t}{\rm e}^{-k^2\Delta^2t}-1 \right) \nonumber \\
= \frac{\eta}{2\sqrt{\pi}\lambda}\int_0^\infty \frac{{\rm d}t}{\sqrt{t}}
{\rm e}^{-\frac{\lambda^2}{4\eta^2 t}} \left[ \theta_3\left({\rm e}^{-t\Delta}\right)
\theta_3\left({\rm e}^{-\frac{t}{\Delta}}\right)-1\right] , \nonumber \\
\label{e2intra}
\end{eqnarray}
where we substituted $t\Delta\to t$ and introduced the Jacobi theta 
function $\theta_3(q,0) \equiv \theta_3(q) = \sum_{j=-\infty}^{\infty} q^{j^2}$
\cite{Gradshteyn}.

The Wigner lattice on the opposite layer at distance $d$ is shifted by 
the vector $({\bm a_1}+{\bm a_2})/2$. 
The square of the distance between the reference particle and the particles 
on the opposite layer becomes 
$r_{jk}^2=(j-1/2)^2 a^2+(k-1/2)^2 a^2\Delta^2+d^2$. 
Proceeding analogously as in the previous case,
we get for the interlayer energy
\begin{equation} \label{e2inter}
E_{\rm inter} =\frac{\eta}{2\sqrt{\pi}\lambda}
\int_0^\infty \frac{{\rm d}t}{\sqrt{t}}
{\rm e}^{-\frac{\lambda^2}{4\eta^2 t}-\eta^2 t}
\theta_2\left({\rm e}^{-t\Delta}\right)\theta_2
\left({\rm e}^{-t/\Delta} \right) ,
\end{equation}
where another Jacobi theta function $\theta_2(q)=\sum_j q^{(j-1/2)^2}$
was introduced.

The total energy per particle $E_{\rm II}$ is a sum $E_{\rm intra}+E_{\rm inter}$.
Using the Poisson summation formula
\begin{equation} \label{poisson}
\sum_{j=-\infty}^\infty {\rm e}^{-(j+\psi)^2 t}
= \sqrt{\frac{\pi}{t}}\sum_{j=-\infty}^{\infty} 
{\rm e}^{2\pi {\rm i} j\psi}{\rm e}^{-(\pi j)^2 /t} ,
\end{equation}
it can be easily shown that in the limit $t\to 0$ the product of 
theta functions $\theta_m({\rm e}^{-t})\theta_m({\rm e}^{-t})\approx \pi/t$ 
for both $m=2,3$.
In the unscreened Coulomb limit $\lambda\to 0$, this would lead to 
the divergence of the corresponding integrals due to the lack of 
the neutralizing background charge.
We ``artificially'' subtract the singular $\pi/t$ terms from the products
of theta functions and simultaneously add the same singular terms
and integrate them explicitly, with the result
\begin{eqnarray}
E_{\rm II} & = & \frac{\eta}{2\sqrt{\pi}\lambda}
\int_0^\infty \frac{{\rm d}t}{\sqrt{t}} {\rm e}^{-\frac{\lambda^2}{4\eta^2 t}}
\bigg\{ \Big[ \theta_3\left({\rm e}^{-t\Delta}\right)
\theta_3\left({\rm e}^{-t/\Delta}\right) \nonumber \\
& & - 1 -\frac{\pi}{t} \Big] +{\rm e}^{-\eta^2 t}
\left[\theta_2\left({\rm e}^{-t\Delta}\right)
\theta_2\left({\rm e}^{-t/\Delta} \right)-\frac{\pi}{t} \right] \bigg\}
\nonumber\\ & & + \pi \frac{\eta^2}{\lambda^2}\left(1+{\rm e}^{-\lambda}\right).
\label{e2tot}
\end{eqnarray}
This corresponds to adding and subtracting the background interaction energy 
\cite{Samaj1}
\begin{equation} \label{eb}
E^{\rm B} = -\pi \frac{\eta^2}{\lambda^2} \left(1+{\rm e}^{-\lambda}\right) .
\end{equation}
The procedure is inevitable in the Coulomb $\lambda\to 0$ limit.
For a positive $\lambda>0$, the procedure is not necessary but it enhances 
substantially the convergence properties of the obtained series.

The integration region $[0,\infty]$ in (\ref{e2intra}) can be split into 
intervals $[0,\pi]$ and $[\pi,\infty]$.
Using the Poisson summation formula (\ref{poisson}), 
the integral over $[\pi,\infty]$ can be rewritten as
\begin{eqnarray}
\int_\pi^\infty \frac{{\rm d}t}{\sqrt{t}} {\rm e}^{-\frac{\lambda^2}{4\eta^2 t}}
\left[ \theta_3\left({\rm e}^{-t\Delta}\right)
\theta_3\left({\rm e}^{-t/\Delta}\right)-1-\frac{\pi}{t}\right] \nonumber\\ 
= \int_\pi^\infty \frac{{\rm d}t}{\sqrt{t}} {\rm e}^{-\frac{\lambda^2}{4\eta^2 t}}
\left( \frac{\pi}{t}\sum_j {\rm e}^{-\frac{(\pi j)^2}{t\Delta}}
\sum_k {\rm e}^{-(\pi k)^2\frac{\Delta}{t}} -1-\frac{\pi}{t}\right) \nonumber\\ 
= \int_0^\pi \frac{\pi\ {\rm d}t'}{t'^{3/2}}
{\rm e}^{-\frac{\lambda^2 t'}{4\eta^2 \pi^2}}
\left( \frac{\pi}{t'}\sum_j {\rm e}^{-\frac{j^2 t'}{\Delta}}
\sum_k {\rm e}^{- k^2 t'\Delta}-1-\frac{t'}{\pi} \right) \nonumber\\ 
= \int_0^\pi \frac{{\rm d}t}{\sqrt{t}} {\rm e}^{-\frac{\lambda^2 t}{4\eta^2 \pi^2}}
\left[\theta_3\left({\rm e}^{-t\Delta}\right)
\theta_3\left({\rm e}^{-t/\Delta}\right)-1-\frac{\pi}{t} \right]. \label{eq11}
\end{eqnarray}
Similarly, 
\begin{eqnarray}
& & \int_\pi^\infty \frac{{\rm d}t}{\sqrt{t}}
{\rm e}^{-\frac{\lambda^2}{4\eta^2 t}}{\rm e}^{-\eta^2 t}
\left[\theta_2\left({\rm e}^{-t\Delta}\right)
\theta_2\left({\rm e}^{-\frac{t}{\Delta}}\right)-\frac{\pi}{t}\right] \nonumber\\
& = & \int_0^\pi\frac{{\rm d}t}{\sqrt{t}}
{\rm e}^{-\frac{\lambda^2 t}{4\eta^2 \pi^2}-\frac{\eta^2 \pi^2}{t}}
\left[\theta_4\left({\rm e}^{-t\Delta}\right) 
\theta_4\left({\rm e}^{-\frac{t}{\Delta}}\right)-1\right], 
\phantom{aaa} \label{eq12}
\end{eqnarray}
where we introduced the Jacobi theta function
$\theta_4(q)=\sum_j (-1)^j q^{j^2}$.

Finally, in close analogy with Ref. \cite{Samaj1} we apply once more 
the Poisson summation formula (\ref{poisson}) for each term 
in the integration from $[0,\pi]$.
The final formula for the energy reads
\begin{widetext}
\begin{eqnarray}
E_{\rm II} & = & \frac{\eta}{2\sqrt{\pi}\lambda}\bigg\{2\sum_{j=1}^\infty\left[
z_{3/2}\left(0,\lambda^2/(4 \pi^2 \eta^2)+j^2\Delta\right)
+z_{3/2}\left(0,\lambda^2/(4 \pi^2 \eta^2)+j^2/\Delta\right)\right]
-\pi z_{1/2}\left(0,\frac{\lambda^2}{4\eta^2\pi^2}\right) \nonumber\\ 
& & + 2\sum_{j=1}^\infty(-1)^j\left[
z_{3/2}\left(\pi^2\eta^2,\lambda^2/(4 \pi^2 \eta^2)+j^2\Delta\right)
+z_{3/2}\left(\pi^2\eta^2,\lambda^2/(4 \pi^2 \eta^2)+j^2/\Delta\right)\right]
-\pi z_{1/2}\left(\lambda^2/(4\eta^2),0\right) \nonumber\\ 
& & + 4\sum_{j,k=1}^\infty (-1)^j(-1)^k 
z_{3/2}\left(\pi^2\eta^2,\lambda^2/(4 \pi^2 \eta^2)+j^2/\Delta+k^2\Delta\right)
+4\sum_{j,k=1}^\infty 
z_{3/2}\left(0,\lambda^2/(4 \pi^2 \eta^2)+j^2/\Delta+k^2\Delta\right) \nonumber\\
& & + 2\sum_{j=1}^\infty(-1)^j\left[
z_{3/2}\left(\pi^2\eta^2,\lambda^2/(4 \pi^2 \eta^2)+j^2\Delta\right)
+z_{3/2}\left(\pi^2\eta^2,\lambda^2/(4 \pi^2 \eta^2)+j^2/\Delta\right)\right]
-\pi z_{1/2}\left(\lambda^2/(4\eta^2),\eta^2\right) \nonumber\\ 
& & + 2\sum_{j=1}^\infty\left[
z_{3/2}\left(\lambda^2/(4\eta^2),j^2\Delta\right)+
z_{3/2}\left(\lambda^2/(4\eta^2),j^2/\Delta\right)\right]
+4\sum_{j,k=1}^\infty z_{3/2}\left(\lambda^2/(4\eta^2),j^2/\Delta+k^2\Delta\right)
\nonumber\\ 
& & + 4\sum_{j,k=1}^\infty z_{3/2}\left[\lambda^2/(4\eta^2),\eta^2+(j-1/2)^2/\Delta
+(k-1/2)^2\Delta\right]\bigg\}
+\pi \frac{\eta^2}{\lambda^2}\left(1+{\rm e}^{-\lambda}\right). \label{e2}
\end{eqnarray}
\end{widetext}
Here, we introduced the function
\begin{equation} \label{znu}
z_{\nu}(x,y)=\int_0^{1/\pi} \frac{{\rm d}t}{t^{\nu}}{\rm e}^{-xt}{\rm e}^{-y/t} .
\end{equation}
It is a generalization of the well-known Misra function \cite{Misra}, 
corresponding to $x=0$, commonly used in lattice summations.
The functions $z_{\nu}(x,y)$ with half-integer values of $\nu$ can be 
expressed in terms of the complementary error function, see Appendix A.
This permits us to use very effectively the MATHEMATICA software and
to derive in Appendix A their asymptotic forms for ($x$ finite, $y\to\infty$)
and ($y$ finite, $x\to\infty$).
The series in the generalized Misra function (\ref{e2}) is quickly converging;
for the known $\lambda=0$ Coulomb cases \cite{Samaj1}, the truncation of 
the series over $j,k$ at $M=1,2,3,4$ reproduces the exact value of the energy 
up to $2,5,10,17$ decimal digits, respectively.
This accuracy even improves itself for $\lambda>0$, so in our numerical 
calculations we keep the truncation of the series at $M=5$. 

The formula (\ref{e2}) is symmetric with respect to the transformation 
$\Delta\to 1/\Delta$.
This symmetry corresponds to an obvious invariance of the energy with respect 
to the lattice rotation around one point by $90$ degrees.

\section{Going from phase I to II} \label{Sec3} 
As was mentioned in Introduction, numerical approaches \cite{Goldoni,Weis} 
predicted that phase I has a region of stability $[0,\tilde{\eta}]$ with 
a very small $\tilde{\eta}>0$ and there is a second-order transition between 
phases I and II.
This small region was expected to vanish ($\tilde{\eta}=0$) in 
the hard-spheres limit $\lambda\to\infty$ \cite{Messina}.
But in the paper \cite{Samaj1} it was shown both analytically and 
numerically that $\tilde{\eta}=0$ in the unscreened Coulomb limit 
$\lambda\to 0$, i.e. phase I exists only for $\eta=0$.
There is no singularity in the ground-state energy, so going from
phase I to phase II is not a phase transition in the usual sense.
In what follows, we derive the same results for any positive $\lambda$.

We know that $\Delta=\sqrt{3}$ for phase I at $\eta=0$.
Let us assume that for $\eta>0$ we have $\Delta=\sqrt{3}-\epsilon$ 
with a small $\epsilon$ and, in close analogy with Ref. \cite{Samaj1}, 
expand the energy (\ref{e2}) in Taylor series: 
\begin{eqnarray} 
E_{\rm II}(\sqrt{3}-\epsilon,\eta,\lambda) & = &
E_{\rm II}(\sqrt{3},\eta,\lambda) + f_1(\eta,\lambda)\epsilon \nonumber \\
& & + f_2(\eta,\lambda)\epsilon^2 + {\cal O}(\epsilon^3) , \label{e21}
\end{eqnarray}
where the expansion functions $f_1(\eta,\lambda)$ and $f_2(\eta,\lambda)$ 
are written explicitly in terms of the generalized Misra functions 
in Appendix B. 
For given $\eta$ and $\lambda$, the extremum of the energy (\ref{e21}) occurs 
at $\epsilon^*$ given by
\begin{equation} \label{eps21eq}
\frac{\partial}{\partial \epsilon} E_{\rm II}(\sqrt{3}-\epsilon,\eta,\lambda)
\Big\vert_{\epsilon=\epsilon^*} \approx f_1(\eta,\lambda) +
2 f_2(\eta,\lambda)\epsilon^* = 0,
\end{equation}
implying
\begin{equation} \label{eps21}
\epsilon^*(\eta,\lambda) \equiv \sqrt{3} - \Delta^*(\eta,\lambda) = 
-\frac{f_1(\eta,\lambda)}{2 f_2(\eta,\lambda)} . 
\end{equation}

For the unscreened Coulomb case $\lambda=0$ it has been shown in 
\cite{Samaj1} that
\begin{equation} \label{Coulombas}
\sqrt{3} - \Delta^*(\eta,0) = - \frac{f_1(\eta,0)}{2 f_2(\eta,0)}  
= 7.14064\ldots \eta^2 + {\cal O}(\eta^4) .  
\end{equation}
This extremum is the minimum of $E_{\rm II}(\epsilon)$. 

\begin{figure}[thb]
\begin{center}
\includegraphics[clip,width=0.42\textwidth]{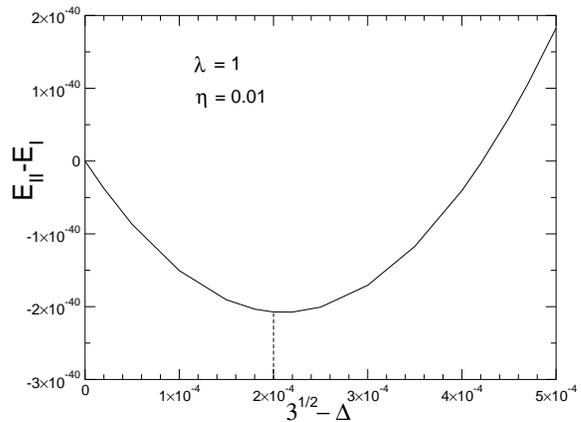}
\caption{$E_{\rm II}(\Delta,\eta)-E_{\rm I}(\eta)$ as a function of 
$\sqrt{3}-\Delta$ for the fixed values of $\lambda=1$ and $\eta=0.01$.
The value of $\epsilon^*=0.0002$, which provides the energy minimum 
according to the asymptotic formula (\ref{extr}) is depicted by the
vertical dashed line for comparison.
Note that the energy differences are extremely small.}
\label{fig:trans12}
\end{center}
\end{figure}

In the case of $\lambda>0$, it is shown in Appendix B that for 
$\eta\ll\lambda$ the coefficient functions can be approximated by
\begin{eqnarray}
f_1(\eta,\lambda) & \approx & -\frac{\eta\lambda}{3^{1/4}\ 4} 
{\rm e}^{-\frac{\lambda}{3^{1/4}\eta}} +
{\cal O}\left(\eta^2 {\rm e}^{-\frac{\lambda}{3^{1/4}\eta}}\right), \nonumber\\
f_2(\eta,\lambda) & \approx & \frac{\lambda}{3^{1/4}\ 16\eta} 
{\rm e}^{-\frac{\lambda}{3^{1/4}\eta}} +
{\cal O}\left(\eta {\rm e}^{-\frac{\lambda}{3^{1/4}\eta}}\right). \label{f1f2}
\end{eqnarray}
The extremum 
\begin{equation} \label{extr}
\sqrt{3} - \Delta^*(\eta,\lambda) 
= - \frac{f_1(\eta,\lambda)}{2 f_2(\eta,\lambda)}  
= 2 \eta^2 + {\cal O}(\eta^4)  
\end{equation}
interestingly does not depend in the leading order on $\lambda$.
It corresponds to the minimum of energy $E_{\rm II}(\epsilon)$ as 
${\partial^2_{\epsilon}} E_{\rm II}(\sqrt{3}-\epsilon,\eta,\lambda)
\vert_{\epsilon=\epsilon^*} = 2 f_2(\eta,\lambda) > 0$.
For $\lambda=1$ and $\eta=0.01$, we checked the result (\ref{extr}) 
numerically in Fig. \ref{fig:trans12}.
One can see that $E_{\rm II}(\epsilon)$, calculated using the complete formula
(\ref{e2}) truncated at $M=5$ has a minimum rather close to the value 
$\epsilon^*=0.0002$ predicted by our asymptotic formula (\ref{eps21}).

We conclude that phase I is stable only at $\eta=0$ and for
an arbitrarily small positive $\eta$ we enter the region of phase II.
It is interesting that the asymptotic $\eta\to 0$ predictions for
the unscreened Coulomb $\lambda=0$ case (\ref{Coulombas}) and for
$\lambda>0$ (\ref{extr}) exhibit the same $\eta^2$ dependence, but
there is a skip in the prefactors from $7.14064\ldots$ at $\lambda=0$
to $2$ for $\lambda>0$. 
The fact that in the previous works \cite{Goldoni,Weis,Messina}
phase I was detected also for small positive values of $\eta$ is
probably related to extremely small deviation of 
$\sqrt{3}-\Delta^*\propto \eta^2$ which are ``invisible'' by
standard numerical methods.

\section{Second-order transition between phases II and III} \label{Sec4}
\begin{figure}[thb]
\begin{center}
\includegraphics[clip,width=0.45\textwidth]{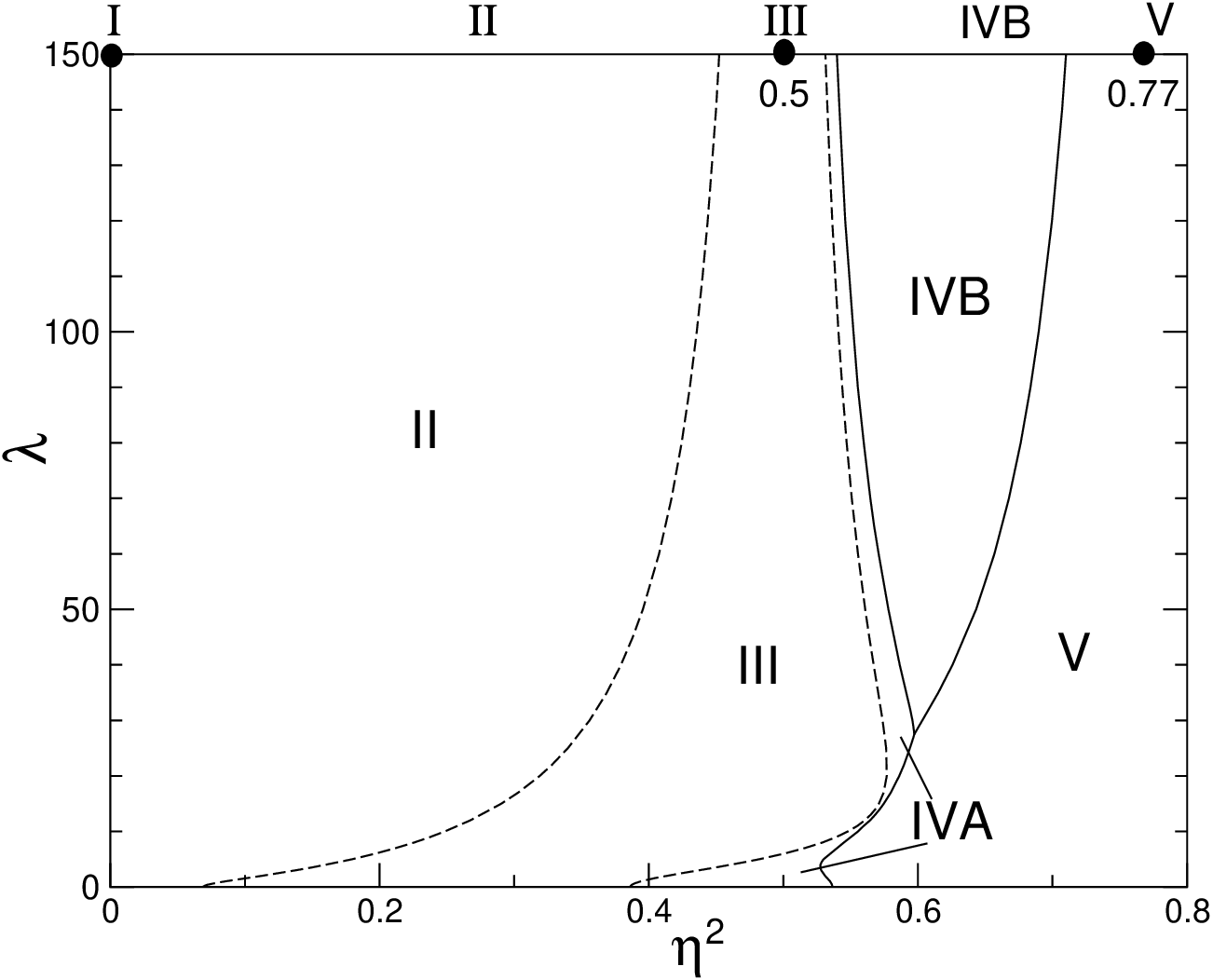}
\caption{Phase diagram of the Yukawa bilayer. Dashed lines denote 
the second-order phase transitions, solid lines correspond 
to the first-order phase transitions.
The important data for the hard-sphere limit $\lambda\to\infty$ 
are added on the top.}
\label{fig:phd}
\end{center}
\end{figure}

Let us parametrize $\Delta=\exp({\epsilon})$. 
The symmetry $\Delta\to 1/\Delta$ of the energy (\ref{e2}) is then 
equivalent to the transformation $\epsilon\to -\epsilon$ and the energy 
is an even function of $\epsilon$.
The Ginsburg-Landau form of its expansion around $\epsilon=0$ reads as
\begin{equation} \label{e2gl}
E_{\rm II}({\rm e}^\epsilon,\eta,\lambda) = E_{\rm III}(\eta,\lambda)
+ g_2(\eta,\lambda)\epsilon^2 + g_4(\eta,\lambda)\epsilon^4 + \ldots
\end{equation}
The explicit expression for $g_2$ is given in Appendix C and 
a rather cumbersome expression for $g_4$ is also at our disposal.
The critical point is given by the vanishing of the prefactor 
\begin{equation} \label{critline}
g_2(\eta^c,\lambda^c) = 0 .
\end{equation}
We used this equation to get the (dashed) critical line between phases II 
and III in Fig. \ref{fig:phd}.
Our definition of $\eta$ differs from that of the dimensionless distance 
in the paper of Messina and L\"owen \cite{Messina}, namely 
$\eta^2 = \eta_{\rm ML}$. 
To maintain the full comparability, we shall present the phase diagram 
using the variable $\eta^2$. 

\subsection{Critical behavior}
\begin{figure}[thb]
\begin{center}
\includegraphics[clip,width=0.44\textwidth]{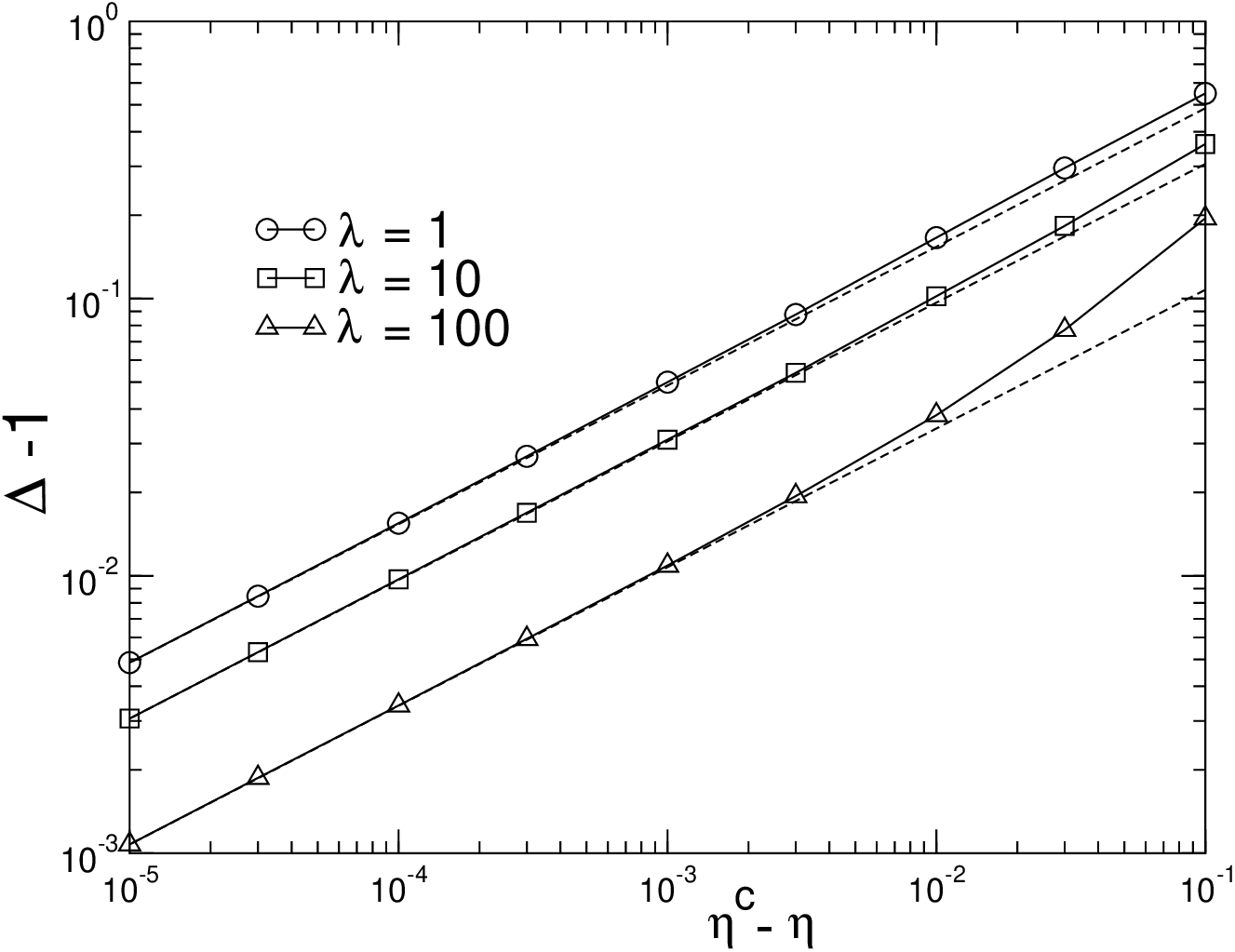}
\caption{Order parameter close to the critical point of the transition II-III 
for three values of $\lambda=1,10,100$. 
Full lines follow from numerical minimization of the energy (\ref{e2}).
The slope of lines is close to $\beta_{\rm MF} = 1/2$.
Dashed lines represent the asymptotic $\eta\to (\eta^c)^-$ 
relation (\ref{epsc}).}
\label{fig:trans23}
\end{center}
\end{figure}

To obtain the critical behavior, we note that the functions $g_2$ and 
$g_4$ in Eq. (\ref{e2gl}) behave in the vicinity of the critical point 
$(\eta^c,\lambda^c)$ as follows
\begin{eqnarray}
g_2(\eta,\lambda) & = & g_{21}(\lambda^c)(\eta^c-\eta)
+ {\cal O}[(\eta^c-\eta)^2], \nonumber \\
g_4(\eta,\lambda) & = & g_{40}(\lambda^c)
+ {\cal O}(\eta^c-\eta), \label{g2g4}
\end{eqnarray}
where $g_{21}(\lambda^c)<0$ and $g_{40}(\lambda^c)>0$ for all $\lambda^c$. 
The minimum energy is reached at $\epsilon^*\approx\Delta^*-1$ given by
\begin{equation} \label{epsceq}
\frac{\partial}{\partial \epsilon} E_{\rm II}({\rm e}^\epsilon,\eta,\lambda)
\big\vert_{\epsilon=\epsilon^*}
\approx 2 g_2(\eta,\lambda)\epsilon^* +4 g_4(\eta,\lambda)(\epsilon^*)^3 = 0 .
\end{equation}
For $\eta>\eta^c$, there is only one solution $\epsilon^*=0$ which
corresponds to the square lattice of phase III.
For $\eta<\eta^c$, we get one trivial (unphysical) solution $\epsilon^* = 0$ 
and two non-trivial conjugate solutions $\pm \epsilon^*$ with
\begin{equation} \label{epsc}
\epsilon^* = \left(-\frac{g_2(\eta,\lambda)}{2 g_4(\eta,\lambda)}\right)^{1/2}
\approx \left(-\frac{g_{21}(\lambda^c)}{2 g_{40}(\lambda^c)}\right)^{1/2}
\sqrt{\eta^c-\eta} , 
\end{equation}
$\eta\to(\eta^c)^-$.
The order parameter $\epsilon^*\propto\sqrt{\eta^c-\eta}$ is thus associated
with the mean-field critical index $\beta_{\rm MF}=1/2$ for every 
$\lambda\ge 0$. 
The dependence of $\Delta-1$ on $\eta^c-\eta$ is shown in 
Fig. \ref{fig:trans23} for three values of $\lambda=1,10,100$. 
Near the critical point ($\eta^c-\eta$ small), the asymptotic
relation (\ref{epsc}) (dashed lines) fits perfectly the numerical data 
from minimization of the energy $E_{\rm II}$ (\ref{e2tot}) (full lines).
In the logarithmic plot, for all values of $\lambda$ the slope of $\Delta-1$ 
vs. $\eta^c-\eta$ is very close to 0.5 in the region of small and 
intermediate values of $\eta^c-\eta$, confirming the value $1/2$ of 
the mean-field critical index $\beta_{\rm MF}$ for all values of $\lambda$. 

From Eq. (\ref{e2gl}), the energy difference of phases
II and III close to the critical point is given by
\begin{equation}
E_{\rm II}({\rm e}^\epsilon,\eta,\lambda) - E_{\rm III}(\eta,\lambda)
\sim - \frac{g_{21}^2(\lambda^c)}{4 g_{40}(\lambda^c)} (\eta^c - \eta)^2 .
\end{equation}
The critical singularity should be of type $(\eta^c - \eta)^{2-\alpha}$
implying the mean-field critical index $\alpha_{\rm MF}=0$ for any $\lambda$.

To obtain another two critical indices, we add to the energy (\ref{e2gl})
the symmetry-breaking term $-h \epsilon$, where a small positive external 
field $h\to 0^+$ is linearly coupled to the order parameter.
The optimization condition for the energy with respect to $\epsilon$
now takes the form
\begin{equation} \label{hcrit}
2 g_2(\eta,\lambda) \epsilon^* + 4 g_4(\eta,\lambda) (\epsilon^*)^3 - h = 0 . 
\end{equation}
At the critical point, since $g_2(\eta^c,\lambda^c)=0$ and 
$g_4(\eta^c,\lambda^c)= g_{40}(\lambda^c)$, we find from (\ref{hcrit}) that
\begin{equation} 
\epsilon^* = \left[ \frac{h}{4 g_{40}(\lambda^c)} \right]^{1/3} .
\end{equation}
This critical singularity should be of type $h^{1/\delta}$,
which leads to the mean-field critical index $\delta_{\rm MF}=3$ 
for any $\lambda$.
Performing the derivative of Eq. (\ref{hcrit}) with respect to $h$,
we find for the field succeptibility close to the critical point:
\begin{equation}
\frac{\partial \epsilon^*}{\partial h} \Bigg\vert_{h=0} =
\frac{1}{- 4 g_{21}(\lambda^c)} \frac{1}{\eta^c-\eta} , \qquad
\eta\to(\eta^c)^- .
\end{equation}
The corresponding critical singularity $(\eta - \eta^c)^{-\gamma}$ leads to
the mean-field critical index $\gamma_{\rm MF}=1$ for arbitrary $\lambda$.

It is easy to verify that our mean-field critical indices 
\begin{equation} \label{MFind}
\alpha_{\rm MF} = 0, \quad \beta_{\rm MF} = \frac{1}{2}, \quad
\gamma_{\rm MF} = 1, \quad \delta_{\rm MF} = 3
\end{equation}
fulfill two standard scaling relations \cite{Ma}
\begin{equation}
2-\alpha = 2\beta + \gamma = \beta (\delta+1) .
\end{equation}
Since there are no fluctuations in our system at zero temperature,
the critical indices $\eta$ and $\nu$, related to the particle
correlation function, are not defined.

\subsection{Coulomb $\lambda\to 0$ limit of the critical line}
\begin{figure}[thb]
\begin{center}
\includegraphics[clip,width=0.44\textwidth]{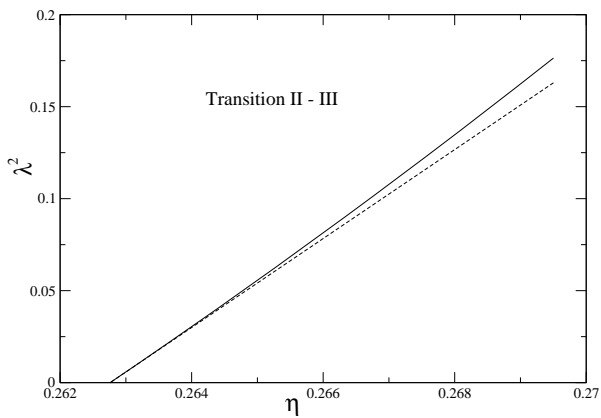}
\caption{The critical line between phases II and III near the unscreened
Coulomb $\lambda\to 0$ limit. 
Full line follows from the numerical evaluation by using
the relation $g_2(\eta,\lambda)=0$.
Dashed line corresponds to the asymptotic formula (\ref{le23}).}
\label{fig:trans23ll}
\end{center}
\end{figure}

We reproduce $\eta^c(0)=0.2627602682$ \cite{Samaj1} 
in the Coulomb $\lambda\to 0$ limit. 
It is shown in Appendix C that the asymptotic $\lambda\to 0$ shape of 
the critical line between phases II and III is parabolic: 
\begin{equation} \label{le23}
\lambda^2 \approx c_{23}[\eta^c-\eta^c(0)] , \qquad c_{23}\approx 24.173744.
\end{equation}
This formula is compared to the critical line evaluated numerically 
by using the relation (\ref{critline}) in Fig. \ref{fig:trans23ll}.

\subsection{Hard-spheres $\lambda\to \infty$ limit of the critical line}
In the hard-spheres limit $\lambda\to\infty$, the critical point for
the ${\rm II}\to {\rm III}$ transition is $(\eta^c)^2\to 1/2$ \cite{Messina}.
The convergence to this value is extraordinarily slow.
Let us analyze this limit in the critical equation $g_2(\eta,\lambda)=0$.
Applying the asymptotic formulas for the generalized Misra functions
(\ref{znuasym}) and (\ref{znuasymp}) to $g_2(\eta,\lambda)$ given by 
the series (\ref{g2}), most summands become exponentially small 
compared to the few leading terms proportional to $\exp(-\lambda/\eta)$.
In particular, we can neglect completely the first four sums in Eq. (\ref{g2}) 
since all terms behave as $\exp(-c\lambda^2)$ and the sixth sum because 
we get at least $\exp(-\sqrt{2}\lambda/\eta)$ for the $j=k=1$ term.
Those leading terms appear in the fifth sum with $j=1$ and the seventh sum 
with $j=k=1$.
The ones with e. g. $j=2$, $k=1$ etc. are exponentially small again compared
to the leading ones.
In the last sum the $z_{7/2}(.,.)$ term has zero prefactor for $j=k$. 
We are left with the three-terms expression
\begin{eqnarray}
g_2(\eta,\lambda) & \approx & z_{7/2}\left(\frac{\lambda^2}{4\eta^2},1\right)
-z_{5/2}\left(\frac{\lambda^2}{4\eta^2},1\right) \nonumber \\ & & 
-\frac{1}{2}z_{5/2}\left(\frac{\lambda^2}{4\eta^2},\eta^2+\frac{1}{2}\right),
\qquad \lambda \gg 1 . \label{g2as0}
\end{eqnarray}
Applying the asymptotic formula (\ref{znuasymp}), we rewrite the rhs of
this expression as
\begin{equation} \label{g2as2}
\frac{\sqrt{\pi}\lambda {\rm e}^{-\frac{\lambda}{\eta}}}{4\eta}
\left[\frac{\lambda}{\eta}+1+\frac{\eta}{\lambda} -
\frac{1+\frac{\eta}{\lambda\sqrt{\eta^2+\frac{1}{2}}}}{\eta^2+\frac{1}{2}}
{\rm e}^{\frac{\lambda}{\eta}\left(1-\sqrt{\eta^2+\frac{1}{2}}\right)}\right].
\end{equation}
The critical condition $g_2(\eta,\lambda) = 0$ implies a transcendental 
formula for $\eta(\lambda)$:
\begin{equation} \label{g2as2root}
\frac{\left(\frac{\lambda}{\eta}+1+\frac{\eta}{\lambda}\right)
\left(\eta^2+\frac{1}{2}\right)}{1+
\frac{\eta}{\lambda\sqrt{\eta^2+\frac{1}{2}}}} =
{\rm e}^{\frac{\lambda}{\eta}\left(1-\sqrt{\eta^2+\frac{1}{2}}\right)} .
\end{equation}
The exponential term can equal to the rational one only if 
$1-\sqrt{\eta^2+1/2}$ is close to zero, i. e. $\eta^2\to 1/2$ 
in the $\lambda\to\infty$ limit as expected.
The next terms of the large-$\lambda$ expansion of $\eta(\lambda)$ 
can be derived straightforwardly, with the result
\begin{equation} \label{e23as}
\eta\approx \frac{1}{\sqrt{2}}-\frac{\ln{\lambda}}{\lambda}
-\frac{\ln{2}}{2\lambda}+{\cal O}\left(\frac{\ln^2{\lambda}}{\lambda^2}\right).
\end{equation}
In general, the series contains the terms of the form
$(\ln \lambda)^m/\lambda^n$ where $m$, $n$ are integers such that
$0\le m\le n$. 
The first correction of type $(\ln\lambda)/\lambda$ explains a slow
convergence of the results as $\lambda\to\infty$.

\begin{figure}[thb]
\begin{center}
\includegraphics[clip,width=0.41\textwidth,height=0.43\textwidth]{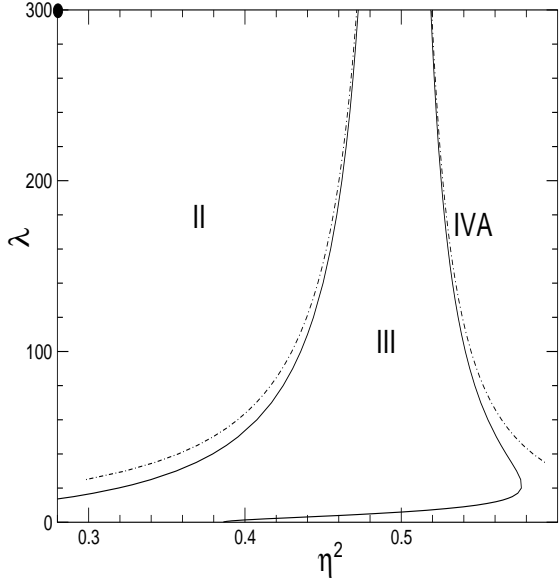}
\caption{An excerption of the phase diagram for Yukawa particles
for the second-order phase transitions ${\rm II}\to {\rm III}$
and ${\rm III}\to {\rm IVA}$. 
The solid lines denote the critical lines obtained numerically by using
Eq. (\ref{critline}) and (\ref{crith}), respectively. 
The dash-dotted lines correspond to the asymptotic large-$\lambda$  
formulas (\ref{e23as}) and (\ref{e34aas}).}
\label{fig:phd2}
\end{center}
\end{figure}

The asymptotic formula (\ref{e23as}), taken for $\eta^2$, is plotted 
in Fig. \ref{fig:phd2} by the dash-dotted line.
We see that it reproduces adequately the numerical results for 
the critical line (solid line) in a large region of the phase diagram.
It can be shown that the next term of the series (\ref{e23as}) reads as
$3 \ln^2\lambda/(2^{3/2}\lambda^2)$; plotting the asymptotic formula 
(\ref{e23as}) with this term included makes the difference with the numerical
solid line invisible by eye.

\section{Energy for structures IVA, IVB and V} \label{Sec5}
It was already mentioned that structures IVA, V and even III are 
special cases of the most general phase IVB.
Hence we will sketch the derivation of the energy per particle for the latter.
The elementary cell is a rhombus with the angle $\phi$ between the vectors 
${\bm a_1}$ and ${\bm a_2}$ of the same magnitude $a$, 
see Fig. \ref{fig:Structures24}.
The density of particles on one plate is $\sigma=1/(a^2\sin\phi)$.
We will prefer the parametrization of the angle by $\delta=\tan (\phi/2)$.
Another free parameter is $\alpha\in [1/3,1/2]$ measuring the diagonal shift 
${\bm c}$ of the lattice on the opposite layer, see formula(\ref{vectorc}).
The square of the lattice vector can be written as 
$\vert{\bm r}_{jk}\vert^2=a^2[(j+k)^2\cos^2(\phi/2)+(j-k)^2\sin^2(\phi/2)]$. 
Next we distinguish the cases when $j+k$ is an even or odd integer and
go to the summation over new indices $m$ and $n$; details of this 
technicality and of the next steps can be found in Sec. III of
paper \cite{Samaj1}.
The main difference is that we get $(n+\alpha)^2$ and $(n-1/2+\alpha)^2$ 
instead of $n^2$ and $(n-1/2)^2$ for the interlayer contribution.
Applying the Poisson formula (\ref{poisson}) creates additional factors 
$\exp(2\pi i n\alpha)$ and $\exp[2\pi i n(\alpha-1/2)]$. 
Reducing the summation over $\{-\infty,\infty\}$ to $\{1,\infty\}$ turns 
these factors to $2\cos(2\pi n\alpha)$ and $2\cos[2\pi n(\alpha-1/2)]$, 
respectively.
The final formula for the energy per particle of phase IVB reads
\begin{widetext}
\begin{eqnarray}
E_{\rm IVB} & = & \frac{\eta}{2\sqrt{2\pi}\lambda}\Bigg(
2\sum_{j=1}^\infty\left[
z_{3/2}\left(0,\frac{\lambda^2}{2 \pi^2 \eta^2}+j^2\delta\right)
+z_{3/2}\left(0,\frac{\lambda^2}{2 \pi^2 \eta^2}+j^2/\delta\right)\right]
\bigg[1+(-1)^j\bigg] \nonumber\\ 
& & + 4\sum_{j,k=1}^\infty \left[ 1+(-1)^{j+k} \right]
z_{3/2}\left(0,\frac{\lambda^2}{2 \pi^2 \eta^2}+j^2/\delta+k^2\delta\right)
-\pi z_{1/2}\left(0,\frac{\lambda^2}{2\eta^2\pi^2}\right) \nonumber\\ 
& & + 2\sum_{j=1}^\infty\left[\cos(2\pi j\alpha)
z_{3/2}\left(\pi^2\eta^2/2,\frac{\lambda^2}{2 \pi^2 \eta^2}+j^2\delta\right)
+z_{3/2}\left(\pi^2\eta^2/2,\frac{\lambda^2}{2 \pi^2 \eta^2}+j^2/\delta\right)
\right] \nonumber\\ & & 
+ 2\sum_{j=1}^\infty\left\{\cos\left[ 2\pi j\left(\alpha-\frac{1}{2}\right)\right]
z_{3/2}\left(\pi^2\eta^2/2,\frac{\lambda^2}{2 \pi^2 \eta^2}+j^2\delta\right)
+(-1)^j z_{3/2}\left(\pi^2\eta^2/2,\frac{\lambda^2}{2 \pi^2 \eta^2}+j^2/\delta
\right)\right\} \nonumber\\ & & 
+ 4\sum_{j,k=1}^\infty\left\{\cos(2\pi j\alpha)
+\cos\left[2\pi j\left(\alpha-\frac{1}{2}\right)\right] (-1)^k\right\}
z_{3/2}\left(\pi^2 \eta^2/2,\frac{\lambda^2}{2 \pi^2 \eta^2}+j^2 \delta
+k^2/\delta\right) \nonumber\\ & & 
+ 2\sum_{j=1}^\infty\left[ z_{3/2}\left(\frac{\lambda^2}{2\eta^2},j^2\delta\right)+
z_{3/2}\left(\frac{\lambda^2}{2\eta^2},j^2/\delta\right)\right]
+4\sum_{j,k=1}^\infty z_{3/2}\left(\frac{\lambda^2}{2\eta^2},j^2/\delta
+k^2\delta\right) -2\pi z_{1/2}\left(\frac{\lambda^2}{2\eta^2},0\right) 
\nonumber\\ & & + \sum_{j,k=-\infty}^\infty 
\left\{z_{3/2}\left[\frac{\lambda^2}{2\eta^2},\eta^2/2+\frac{1}{\delta}
(j+\alpha)^2+k^2\delta\right] + z_{3/2}\left[\frac{\lambda^2}{2\eta^2},
\eta^2/2+\frac{1}{\delta}(j+\alpha-1/2)^2+(k-1/2)^2\delta\right]\right\}
\nonumber\\ & & - 2\pi z_{1/2}\left(\frac{\lambda^2}{2\eta^2},\eta^2/2\right)
+ 4 \sum_{j,k=1}^\infty z_{3/2}\left[\frac{\lambda^2}{2\eta^2},
\frac{1}{\delta}\left(j-\frac{1}{2}\right)^2
+\left(k-\frac{1}{2}\right)^2\delta\right]\Bigg)
 +\pi \frac{\eta^2}{\lambda^2}\left(1+{\rm e}^{-\lambda}\right). \label{e4b}
\end{eqnarray}
\end{widetext}

\section{Transition between phases III and IVA} \label{Sec6}
One can verify that for the structure IVA with $\alpha=1/2$ 
the energy (\ref{e4b}) possesses the symmetry $\delta\to 1/\delta$. 
The case $\delta=1$ or $\phi=\pi/2$ is the fixed point of the transformation
$\delta\to 1/\delta$ and corresponds to the critical point between
phases III and IVA.
In full analogy with the transition between phases II and III, 
we parametrize $\delta=\exp(-\epsilon)$ so that the energy of phase IVA 
becomes an even function of $\epsilon$.
The expansion of the energy (\ref{e4b}) around the critical point $\delta=1$ 
in powers of small $\epsilon$ takes the form 
\begin{eqnarray} 
E_{\rm IVA}({\rm e}^{-\epsilon},\eta,\lambda) & = & E_{\rm III}(1,\eta,\lambda)
+ h_2(\eta,\lambda) \epsilon^2 \nonumber \\
& & + h_4(\eta,\lambda)\epsilon^4 + \ldots . \label{e4agl}
\end{eqnarray}
The explicit formula for $h_2$ in terms of the generalized Misra functions
is presented in Appendix D and $h_4$ is also at our disposal.
The critical line between phases III and IVA is once again given by 
vanishing of the prefactor 
\begin{equation} \label{crith}
h_2(\eta^c,\lambda^c) = 0 ,
\end{equation} 
see Figs. \ref{fig:phd} and \ref{fig:phd2}.

\subsection{Critical behavior}
The expansion of the coefficients $h_2$ and $h_4$ around the critical point 
$(\eta^c,\lambda^c)$ is analogous to the previous case of the
second-order transition between phases II and III. 
The leading terms are $h_2(\eta,\lambda)\approx h_{21}(\lambda^c)(\eta-\eta^c)$ 
and $h_4(\eta,\lambda)\approx h_{40}(\lambda^c)$, where 
$h_{21}(\lambda^c)<0$ and $h_{40}(\lambda^c)>0$ for all $\lambda_c$.
Optimizing the energy $E_{\rm IVA}$ with respect to $\epsilon$, 
the stationary solution $\epsilon^* = 1-\delta^*$ behaves as
\begin{equation} \label{epsc34}
\epsilon^*=\left(-\frac{h_2(\eta,\lambda)}{2 h_4(\eta,\lambda)}\right)^{1/2}
\approx \left(-\frac{h_{21}(\lambda^c)}{2 h_{40}(\lambda^c)}\right)^{1/2}
\sqrt{\eta - \eta^c}
\end{equation}
with $\eta\to (\eta^c)^+$.
The order parameter $\epsilon^*$ has again the singular behavior 
of mean-field type with critical index $\beta_{\rm MF} = 1/2$.
We tested this results numerically in a plot analogous to 
Fig. \ref{fig:trans23} and got the slope $\beta\approx 0.499$.
Without going into details, also other critical indices attain their 
mean-field values (\ref{MFind}).

\subsection{Coulomb $\lambda\to 0$ limit of the critical line}
The week screening (small $\lambda$) case of the phase transitions III-IVA 
and IVA-V was studied by Monte Carlo methods in Ref. \cite{Mazars08}.

\begin{figure}[thb]
\begin{center}
\includegraphics[clip,width=0.44\textwidth]{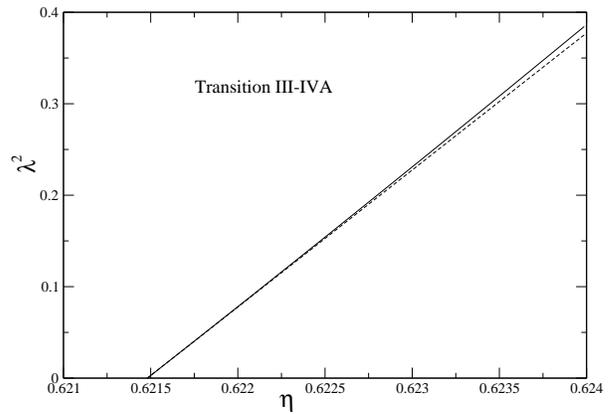}
\caption{Transition III-IVA near the Coulomb $\lambda\to 0$ limit. 
The full line follows from the numerical treatment of 
the relation $h_2(\eta,\lambda)=0$.
The dashed line corresponds to the asymptotic formula (\ref{le34a}).}
\label{fig:trans34all}
\end{center}
\end{figure}

We reproduce $\eta^c(0) = 0.6214809246$ \cite{Samaj1} in the Coulomb 
$\lambda\to 0$ limit. 
The asymptotic shape of the critical line for small $\lambda$ is again 
parabolic, see Appendix D:
\begin{equation} \label{le34a}
\lambda^2\approx c_{34}[\eta^c-\eta^c(0)],\qquad c_{34}\approx 149.7837254 .
\end{equation}
This asymptotic result is compared with the numerical calculation of
the critical line directly from the relation (\ref{crith}) 
in Fig. \ref{fig:trans34all}.

\subsection{Hard-spheres $\lambda\to \infty$ limit of the critical line}
In the hard-spheres limit $\lambda\to\infty$, the critical point
for the ${\rm III}\to {\rm IVA}$ transition is $(\eta^c)^2\to 1/2$ 
\cite{Messina}, the same as in the previous case of 
the ${\rm II}\to {\rm III}$ transition.
Let us analyze the large-$\lambda$ limit of the critical relation 
$h_2(\eta,\lambda)=0$.
In the same way as for $g_2$, we get three leading terms from the seventh 
and ninth (last) sums of Eq. (\ref{h2}):
\begin{eqnarray}
\frac{1}{16} z_{7/2}\left(\frac{\lambda^2}{2\eta^2},\frac{\eta^2}{2}
+\frac{1}{4}\right)
-\frac{1}{4}z_{5/2}\left(\frac{\lambda^2}{2\eta^2},\frac{\eta^2}{2}
+\frac{1}{4}\right) \nonumber\\
-\frac{1}{2}z_{5/2}\left(\frac{\lambda^2}{2\eta^2},\frac{1}{2}\right) = 0,
\qquad \lambda \gg 1 . \label{h2as0}
\end{eqnarray}
The application of the asymptotic relations (\ref{znuasymp}) 
to this equation implies
\begin{eqnarray}
\sqrt{\frac{\pi}{2}}\frac{\lambda}{\eta}{\rm e}^{-\frac{\lambda}{\eta}}
\Bigg\{\Bigg[\frac{\lambda}{8\eta \left(\eta^2+\frac{1}{2}\right)^{3/2}}
\Bigg(1+\frac{3\eta}{\lambda\sqrt{\eta^2+\frac{1}{2}}} \nonumber\\ 
+\frac{3\eta^2}{\lambda^2(\eta^2+\frac{1}{2})}\Bigg)
-\frac{1}{2(\eta^2+\frac{1}{2})}
\left(1+\frac{\eta}{\lambda\sqrt{\eta^2+\frac{1}{2}}}\right) \Bigg] \nonumber\\ 
\times {\rm e}^{\frac{\lambda}{\eta}\left(1-\sqrt{\eta^2+1/2}\right)}
-\left(1+\frac{\eta}{\lambda}\right) \Bigg\} = 0 . \label{h2as2}
\end{eqnarray}
The root of the expression in the largest parentheses yields
\begin{eqnarray}
\eta & \approx & \frac{1}{\sqrt{2}} + \frac{\ln \lambda}{\lambda} +
- \frac{5\ln 2}{2\lambda} \nonumber \\ & & +
\frac{3}{4} \sqrt{2}\left(\frac{\ln{\lambda}}{\lambda}\right)^2+
{\cal O}\left(\frac{\ln{\lambda}}{\lambda^2}\right) . \label{e34aas}
\end{eqnarray}
This asymptotic formula, taken for $\eta^2$, is plotted 
in Fig. \ref{fig:phd2} by the dash-dotted line.
The comparison with the numerical results for the critical line (solid line) 
is very good.

\section{Phase transitions IVA - V, IVA - IVB and IVB - V} \label{Sec7}
All phase transitions IVA - V, IVA - IVB and IVB - V are of first order
due to a discontinuous change of both structure parameters 
$\delta$ and $\alpha$.
For phase IVB one has to minimize numerically the energy (\ref{e4b}) 
with respect to two parameters $\delta$ and $\alpha$, which is tedious 
but feasible.

\begin{figure}[thb]
\begin{center}
\includegraphics[clip,width=0.41\textwidth]{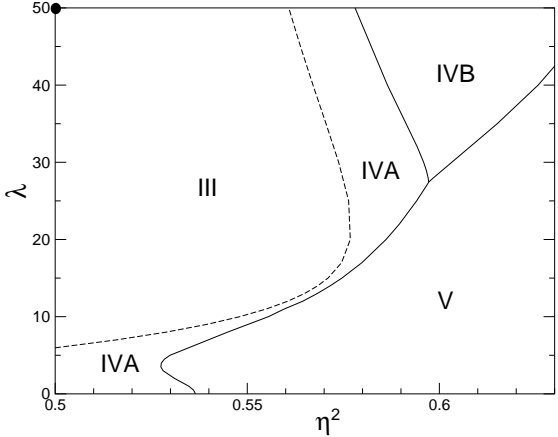}
\caption{A detailed view of the phase diagram around the tricritical point.
The sector of phase IVA is connected by a narrow channel.}
\label{fig:phd3}
\end{center}
\end{figure}

We found that for $\lambda < 27.4436$ phase IVA goes over directly to 
phase V without entering the intermediate phase IVB.
On the transition line, the parameter $\alpha$ jumps from 1/2 to 1/3.
In the Coulomb limit $\lambda\to 0$ we get $\eta^t(0)=0.732416$,
$\delta^t=0.69334$ for phase IVA \cite{Samaj1} whereas for the rigid phase V 
$\delta^t=\tan(\pi/6)=1/\sqrt{3}\approx 0.57735$.
The shape of the transition line is again parabolic, we can approximate
it empirically by 
\begin{equation}
\lambda^2\approx c_{4A5} [\eta^t(0)-\eta^t] , \qquad c_{4A5}\approx 805.3 , 
\end{equation}
but now the parabola is reversed giving rise to the multiple reentrant
behavior, see Figs. \ref{fig:phd} and \ref{fig:phd3}.

The non-trivial $\delta^t$ for phase IVA increases to approximately 0.926 
at $\lambda = 14$ and then slightly decreases to 0.853081 at $\lambda=27.4436$.
In Ref. \cite{Messina} it was anticipated that there exist two disjunct 
regions of phase IVA in the phase diagram. 
Our more precise calculations indicate that there exists a narrow connecting 
channel merging these two regions into one, see Fig. \ref{fig:phd3}.
The maximum value of $\delta^t$ for phase IVA is achieved when the channel 
is the most narrow so that it does not decrease too much from 
the value $\delta=1$ for phase III.
Looking at Figs. \ref{fig:phd} and \ref{fig:phd3} we can confirm 
the double reentrant scenario IVA-V-IVA-III-IVA-IVB \cite{Messina}, 
restricted to a more precise interval $0.5275<\eta^2<0.53643$.

For $\lambda>27.4436$, the phase IVB takes place and we have first-order 
transitions IVA-IVB and IVB-V, see Figs. \ref{fig:phd} and \ref{fig:phd3}.

As concerns the transition line IVA - IVB, for $\lambda\to\infty$ 
it should asymptotically approach the value $(\eta^t)^2\to 1/2$ so that 
phase IVA is absent in the hard-spheres limit \cite{Messina}.
The value of $\delta^t$ in phase IVA increases from 0.85308 
at $\lambda=27.4436$ towards 1 for very large $\lambda$.
Concerning phase IVB, $\delta^t$ increases from 0.763284 at $\lambda=27.4436$
to 1 for very large $\lambda$ and the other parameter $\alpha^t$ from 
0.41358 to 0.5 along the same transition line.
Thus, in the hard-spheres limit, the values $\delta^t\to 1$ and
$\alpha^t\to 1/2$ of phase III (see the top of Fig. \ref{fig:phd})
will be attained as expected.

Still in the hard-spheres limit $\lambda\to\infty$, the transition line 
IVB-V should reach the point $\eta^t \approx 0.877\ldots$ \cite{Messina}. 
Numerically, we got mere $\eta^t = 0.864133\ldots$ even for $\lambda = 500$.
The convergence is rather slow again, we have $\delta^t=0.58102$ 
and $\alpha^t=0.334428$ for phase IVB at the same $\lambda=500$ value, 
gradually approaching the values 0.57735 and 1/3 of phase V, respectively,
with ${\cal O}(1/\lambda)$ corrections of both structure parameters.
Now we want to derive the above hard-spheres result from our formalism.
We recall that the energy $E_{\rm IVB}$ is given by Eq. (\ref{e4b}) and 
$E_{\rm V}$ is its special case for $\delta=1/\sqrt{3}$ and $\alpha=1/3$. 
We apply the asymptotic formulas (\ref{znuasym}) and (\ref{znuasymp}) 
to the $\lambda\to\infty$ limit of Eq. (\ref{e4b}) and neglect exponentially 
small terms.
Five summands remain dominant; one from the sixth sum with $j=1$, three from 
the eighths (last but one) sum, namely both terms with $j=k=0$ and the second 
one with $j=0$, $k=1$ plus the $j=k=0$ term from the ninth (last) sum:
\begin{eqnarray}
E_{\rm IVB} & \approx & \frac{\eta}{2\sqrt{2\pi}\lambda}
\biggl\{2z_{3/2}\left(\frac{\lambda^2}{2\eta^2},\delta\right) \nonumber\\
& & +z_{3/2}\left(\frac{\lambda^2}{2\eta^2},
\frac{\eta^2}{2}+\frac{\alpha^2}{\delta}\right) \nonumber\\ 
& & + 2\ z_{3/2}\left[\frac{\lambda^2}{2\eta^2},
\frac{\eta^2}{2}+\frac{(\alpha-1/2)^2}{\delta}+\frac{\delta}{4}\right]
\nonumber\\ 
& & +4\ z_{3/2}\left(\frac{\lambda^2}{2\eta^2},
\frac{1}{4\delta}+\frac{\delta}{4}\right) \biggr\} . \label{eivbas}
\end{eqnarray}
Notice that two identical terms merged to the one on the third line. 
All these summands should be of the same order for very large $\lambda$.
Since the asymptotic relations (\ref{znuasymp}) imply that
$z_{\nu}(x,y)\propto \exp(-2\sqrt{x y})$ for $x\to\infty$, and the first 
argument $x=\lambda^2/(2\eta^2)$ is common for the summands, the second 
arguments must coincide as well.
Thus we have
\begin{equation} \label{eta2hs}
\delta = \frac{\eta^2}{2}+\frac{\alpha^2}{\delta} = \frac{\eta^2}{2}
+\frac{(\alpha-\frac{1}{2})^2}{\delta}+\frac{\delta}{4}
= \frac{1}{4\delta}+\frac{\delta}{4} . 
\end{equation}
This equalities yield the expected asymptotic values of the structure
parameters $\delta = 1/\sqrt{3}$ and $\alpha = 1/3$.
Simultaneously,
\begin{equation} \label{etat-hs}
\eta^t = \frac{2}{3^{3/4}}\approx 0.877383\ldots ,\qquad \lambda\to\infty.
\end{equation}
This value can be rederived from purely geometric considerations, too. 
We have already mentioned that the particle density at one plate in phase V 
is $\sigma=1/[a^2\sin(\pi/3)]=2/(a^2 \sqrt{3})$.
For dense packed hard spheres of radius $a$, the perpendicular distance of 
two layers of triangular lattices is $d=\sqrt{2/3}\ a$, 
see e.g. \cite{Schmidt}.
Inserting these values into $\eta = d \sqrt{\sigma}$ yields immediately 
(\ref{etat-hs}). 

We confirm that in the hard-spheres limit the transition IVB-V will undergo 
no stepwise changes of structure parameters and it will be of the second order,
as expected.

\begin{figure}[thb]
\begin{center}
\includegraphics[clip,width=0.45\textwidth]{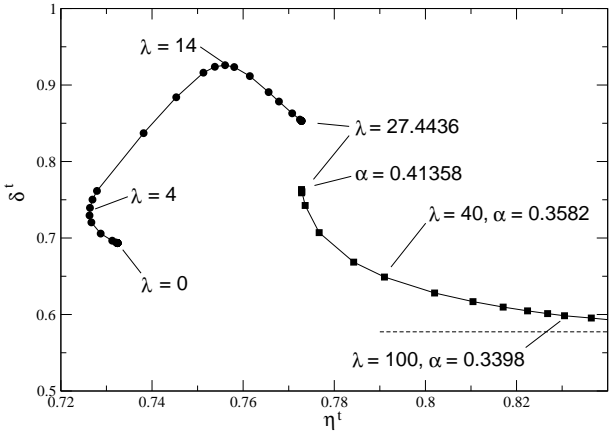}
\caption{The transition parameters $\delta^t$ and $\eta^t$ along 
the phase transition lines IVA-V (left) and IVB-V (right). 
Along two line fragments, $\lambda$ increases from 0 to $\infty$.
For the left fragment, the parameter $\alpha=1/2$ for phase IVA and 
$\alpha=1/3$ for phase V. 
There is a discontinuity in the parameters $\delta$ and $\alpha$ 
between phases IVA and IVB at the tricritical point with $\lambda=27.4436$.
The values of $\alpha$ on the right fragment correspond to phase IVB.
In the hard-spheres limit $\lambda\to\infty$, the line ends up at 
the critical point $(\eta^c=0.877383,\delta^c=1/\sqrt{3})$.}
\label{fig:deltaeta}
\end{center}
\end{figure}

In Fig. \ref{fig:deltaeta}, we present the transition values of the structure 
parameters $\delta^t$ and $\alpha^t$ for phases IVA and IVB at first-order 
transitions IVA - V (left, $\lambda\in [0,27.4436]$) 
and IVB - V (right, $\lambda\in [27.4436,\infty]$).
The left and right line fragments are separated by a gap, illustrating 
the step-wise change of structure parameters when going from phase IVA to IVB.
We recall that the parameters of phase V are always fixed to 
$\delta^t=1/\sqrt{3}$ and $\alpha^t=1/3$. 
We found a tricritical point at $\eta^c = 0.772814$ and $\lambda^c=27.4436$
where the three phases IVA, IVB and V coexist.

\section{The energy plot} \label{Sec8}
\begin{figure}[thb]
\begin{center}
\includegraphics[clip,width=0.45\textwidth]{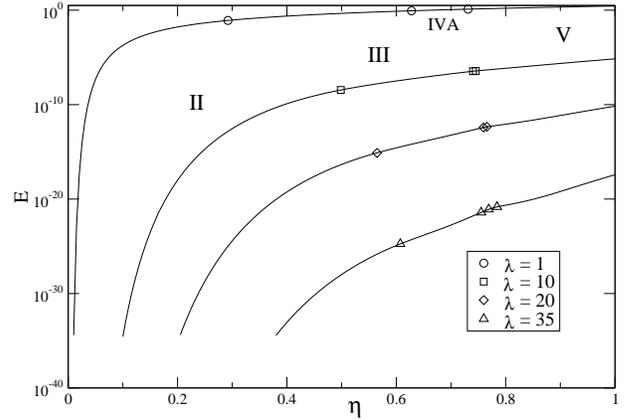}
\caption{The dependence of the energy per Yukawa particle $E$ on the 
dimensionless distance $\eta$ for four values of $\lambda=1,10,20,35$, 
in semilogarithmic scale.}
\label{fig:energy-eta}
\end{center}
\end{figure}

We want to compare the values of the optimized energy per particle
for various values of $\lambda$ and $\eta$.
We plot $E(\eta)$ for several fixed values of $\lambda$ 
in Fig. \ref{fig:energy-eta}.
These energies vary by orders of magnitude, thus we have chosen 
semilogarithmic scale.

First we consider two limiting cases.
For $\eta\ll 1$ and $\lambda>0$, according to (\ref{e21}) and (\ref{f1f2}) 
the energy of the corresponding phase II 
$\ln(E_{\rm II})\approx -3^{1/4}\lambda/\eta$ 
and so $\ln {E}$ diverges if $\eta\to 0$.
More interesting is the optimal energy of phase V, $E_{\rm V}$, for $\eta\gg 1$. 
We were used to get the Coulomb limit as $\lambda\to 0$, but we can obtain 
this limit also for medium $\lambda$ and very large $\eta$, as the ratio 
$\lambda/\eta\to 0$ again.
For $\eta\gg \lambda$, using the asymptotic formulas for the generalized
Misra functions (Appendix A) we obtain from (\ref{e4b}) that
\begin{equation} \label{e5as}
E_{\rm V} = \pi \frac{\eta^2}{\lambda^2}\left(1+{\rm e}^{-\lambda}\right)
+ c_M\frac{\eta}{\lambda} +{\cal O}(1) ,
\end{equation}
where $c_M=-1.9605158...$ is the Madelung constant of the Coulomb potential
for the hexagonal lattice; for an explicit representation of the Madelung
constant in terms of $z_{\nu}(0,y)$ functions, see Eq. (24) with 
$\Delta=\sqrt{3}$ and $\eta=0$ of Ref. \cite{Samaj1}.
The leading term is the (minus) background energy (\ref{eb}). 

\begin{figure}[thb]
\begin{center}
\includegraphics[clip,width=0.45\textwidth]{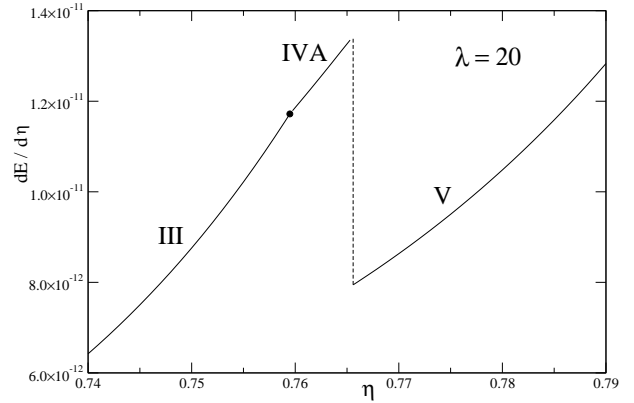}
\caption{The derivative $\partial E/\partial \eta$ for $\lambda=20$. 
Full circle corresponds to the second-order transition III-IVA, 
dashed line marks the discontinuity at the first-order transition IVA-V.}
\label{fig:de-eta}
\end{center}
\end{figure}

We see in Fig. \ref{fig:energy-eta} for few fixed values of $\lambda$ that 
the energy is a monotonously increasing function of the dimensionless
distance $\eta$.
This means that the force between the plates is always attractive.
The non-analyticities at transition points are not clearly manifested 
in this scale.
Therefore, for $\lambda=20$, we performed the derivative 
$\partial E/\partial \eta$, directly for rigid structures and 
numerically using $E_{\rm IVA}$ minimized with respect to $\delta$ 
for phase IVA. 
The obtained results are plotted in Fig. \ref{fig:de-eta}. 
We see the expected continuous but non-analytic behavior at 
the second-order transition point III-IVA as well as 
a jump discontinuity at the first order transition IVA-V.

\section{Internal parameters of the phases near hard-spheres limit} \label{Sec9}
In and close to the limit of hard spheres $\lambda\to\infty$, the expressions
for the energies of the structures in terms of the generalized Misra 
functions admit an asymptotic analysis.
This fact permits us to determine the $\eta$-dependence of the structure 
parameters of the present soft phases II and IVB in the $\lambda\to\infty$ 
limit and eventually to derive their leading correction for large but 
finite $\lambda$.

\subsection{Aspect ratio $\Delta$ of phase II at and near hard spheres}
The dependence of the aspect ratio $\Delta_{\rm HS}$ on $\eta$ for phase II 
is well known in the hard-spheres limit $\lambda\to\infty$ \cite{Messina}:
\begin{equation} \label{DeltaHS}
\Delta_{\rm HS}(\eta) = \sqrt{4\eta^4+3} - 2\eta^2 .
\end{equation}
In the following, we derive this result and the first $1/\lambda$ 
correction to it by using our method.

\begin{figure}[thb]
\begin{center}
\includegraphics[clip,width=0.45\textwidth]{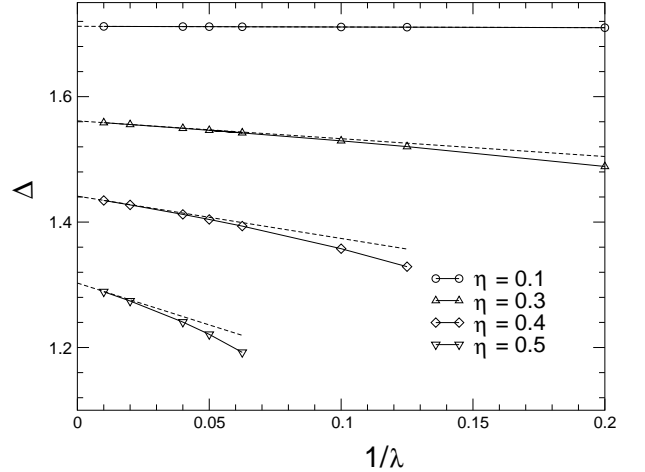}
\caption{The aspect ratio $\Delta$ of phase II vs. $\lambda$ for
four values of $\eta=0.1,0.3,0.4,0.5$. 
The solid lines correspond to numerical calculations.
The asymptotic $\lambda\to\infty$ result (\ref{delta}), (\ref{aeta})
is represented by dashed lines.}
\label{fig:del-lam}
\end{center}
\end{figure}

For $\lambda\gg 1$, most of terms in the energy of phase II (\ref{e2}) 
become exponentially small (we exclude from the discussion trivial terms 
which do not depend on $\Delta$); only the term $j=1$ in the sixth sum 
and the term $j=k=1$ in the eighth (last) sum contribute. 
As soon as $\Delta>1$, using (\ref{znuasymp}) we get
\begin{eqnarray} 
E_{\rm II} \approx \frac{\eta}{\sqrt{\pi}\lambda}
\left[z_{\frac{3}{2}}\left(\frac{\lambda^2}{4 \eta^2},\frac{1}{\Delta}\right)
+2 z_{\frac{3}{2}}\left(\frac{\lambda^2}{4 \eta^2},\eta^2
+\frac{\Delta}{4}+\frac{1}{4\Delta}\right)\right] \nonumber \\
\approx \frac{\eta}{\lambda}
\left(\sqrt{\Delta} {\rm e}^{-\frac{\lambda}{\eta\sqrt{\Delta}}}
+\frac{2}{\sqrt{\eta^2+\frac{\Delta}{4}+\frac{1}{4\Delta}}}
{\rm e}^{-\frac{\lambda}{\eta}\sqrt{\eta^2+\frac{\Delta}{4}+\frac{1}{4\Delta}}}
\right) . \nonumber \\ \label{e2hs}
\end{eqnarray}
The minimum of the energy is given by $\partial E_{\rm II} /\partial \Delta =0$, 
which implies
\begin{eqnarray}
\left( \frac{1}{2\sqrt{\Delta}}+\frac{\lambda}{2\Delta\eta}\right)
{\rm e}^{-\frac{\lambda}{\eta\sqrt{\Delta}}} =
\frac{1-\frac{1}{\Delta^2}}{4 \eta^2+\Delta+\frac{1}{\Delta}}
\nonumber \\ \times \left(
\frac{1}{\sqrt{\eta^2+\frac{\Delta}{4}+\frac{1}{4\Delta}}}+\frac{\lambda}{\eta}
\right)
{\rm e}^{-\frac{\lambda}{\eta}\sqrt{\eta^2+\frac{\Delta}{4}+\frac{1}{4\Delta}}}. \label{del2}
\end{eqnarray}
If we want to reproduce just the hard-spheres limit $\lambda\to\infty$, 
we can say that exponentials are by far more significant than rationals 
and their arguments must become the same, i.e. 
$\sqrt{\eta^2+\Delta/4+1/(4\Delta)}=1/\sqrt{\Delta}$ which
leads to the known result (\ref{DeltaHS}).
Numerics suggests that the correction is of the type $1/\lambda$, i. e. 
\begin{equation} \label{delta}
\Delta \approx \Delta_{\rm HS} + \frac{a(\eta)}{\lambda}
=\sqrt{4\eta^4+3}-2\eta^2 + \frac{a(\eta)}{\lambda} .
\end{equation}
We put the exponentials on one side, insert (\ref{delta}) and expand 
$\sqrt{\eta^2+\Delta/4+1/(4\Delta)}-1/\sqrt{\Delta}$ up to the order 
$1/\lambda$. 
The absolute term vanishes and we have
\begin{eqnarray} 
\exp{\left[{-\frac{a(\eta)}{4\eta}\ \frac{3+4\eta^4-2\eta^2\sqrt{4\eta^4+3}}
{(\sqrt{4\eta^4+3}-2\eta^2)^{3/2}}}\right]} = \nonumber \\  
\frac{\eta^2+\frac{\Delta}{4}+\frac{1}{4\Delta}}{2\Delta
\left(\frac{1}{4}-\frac{1}{4\Delta^2}\right)} 
\approx \frac{1}{1+4\eta^4-2\eta^2\sqrt{4\eta^4+3}}, \label{del3}
\end{eqnarray}
where we considered $\Delta\approx\Delta_{\rm HS}$ on the second line.
From this relation we readily get 
\begin{eqnarray}
a(\eta) & = & \frac{\left(\sqrt{4\eta^4+3}-2\eta^2\right)^{3/2}4\eta}{3
+4\eta^4-2\eta^2\sqrt{4\eta^4+3}} \nonumber \\ & &
\times\ln{\left[1+4\eta^4-2\eta^2\sqrt{4\eta^4+3}\right]}. \label{aeta}
\end{eqnarray}
The value of $a(\eta)$ is negative in the whole interval $0<\eta<1/\sqrt{2}$
of phase II.
We find $a(\eta)\approx-8\times 3^{1/4}\eta^3$ for $\eta\ll 1$, 
confirming once more that phase II is entered directly from phase I
for any small positive $\eta$.
We tested the asymptotic result (\ref{delta}), (\ref{aeta})
numerically, see Fig. \ref{fig:del-lam}.

\subsection{Parameters $\delta$ and $\alpha$ of phase IVB in 
the hard-spheres limit}
As was shown above, in the hard-spheres limit $\lambda\to\infty$ phase IVB 
takes place in the interval $\eta\in [1/\sqrt{2},2/3^{3/4}]$.
Let us study the $\lambda\to\infty$ limit of the energy $E_{\rm IVB}$ 
(\ref{e4b}); terms which do not depend on $\delta$ or $\alpha$ are 
automatically excluded from the discussion.
For very large $\lambda$ and general $\eta$ only the last two sums contribute, 
the remaining sums are exponentially small.
The eighth sum has three important terms - one from the first $z_{3/2}$ summand 
with $j=k=0$ and two identical ones from the second $z_{3/2}$ summand 
with $j=k=0$ and $j=0,k=1$.
From the ninth (last) sum we take the $j=k=0$ term.
The result is
\begin{eqnarray}
E_{\rm IVB} & \approx & \frac{\eta}{2\sqrt{2\pi}\lambda}\Bigg[
z_{3/2}\left(\frac{\lambda^2}{2\eta^2},\eta^2/2+\frac{\alpha^2}{\delta}\right)
\nonumber\\
&+&2z_{3/2}\left(\frac{\lambda^2}{2\eta^2},\eta^2/2+\frac{(\alpha-1/2)^2}{\delta}
+\frac{\delta}{4}\right)
\nonumber\\
&+&4z_{3/2}\left(\frac{\lambda^2}{2\eta^2},\frac{1}{4\delta}
+\frac{\delta}{4}\right) \Bigg]. \label{e4bapp}
\end{eqnarray}
We notice that two more terms can become important in special limits.
The first is the $j=-1,\ k=0$ term from the eighth sum, 
the first $z_{3/2}$ summand, which contributes only in the limit $\delta\to 1$
(i. e. $\eta^2\to 1/2$).  
The other possibly important term can be found in Eq. (\ref{eivbas}) 
as the first one in the bracket, but it plays role only if 
$\eta^2\to 4/(3\sqrt{3})$ and it can be omitted for general $\eta$ as well.

Now we apply the asymptotic relations (\ref{znuasymp}) to 
the energy (\ref{e4bapp}). 
The optimization of the energy with respect to parameters $\delta$ and
$\alpha$ leads to the equations $\partial E/\partial \delta = 0$ and 
$\partial E/\partial \alpha = 0$. 
What we get are certain products of rational functions and exponentials. 
To have a non-trivial solution in the $\lambda\to\infty$ limit, 
the dominant exponentials must have the same arguments which yields   
\begin{equation} \label{exparg}
\eta^2/2+\frac{\alpha^2}{\delta} =\eta^2/2+\frac{(\alpha-1/2)^2}{\delta}
+\frac{\delta}{4} =\frac{1}{4\delta}+\frac{\delta}{4} .
\end{equation}
This set set of equations can be readily rewritten as 
\begin{equation} \label{deltaalpha}
\alpha = \frac{1}{4}\left(1+\delta^2\right), \qquad
\alpha^2 -\alpha+\frac{\eta^2}{2}\delta = 0.
\end{equation}
The quartic equation for $\delta$ follows
\begin{equation} \label{delta4}
\delta^4-2\delta^2+8\eta^2\delta-3 = 0 .
\end{equation}
The discriminant of this equation $-2^{12}(3-14\eta^4+27\eta^8)$ is negative
for any $\eta$. 
Consequently, we get two complex roots and two real ones.
It turns out that one of the real roots is negative and the only
physical - real positive - root is given by Cardano formulas as follows
\begin{equation} \label{delta2}
\delta = -S(\eta)+\frac{1}{2}\sqrt{-4S^2(\eta)+4+\frac{8\eta^2}{S(\eta)}},
\end{equation}
where
\begin{equation} \label{S}
S(\eta) = \frac{1}{2}\sqrt{\frac{4}{3}+\frac{1}{3}
\left[Q(\eta)-\frac{32}{Q(\eta)}\right]}
\end{equation}
with
\begin{equation} \label{Q}
Q(\eta)=2^{5/3}\left(27\eta^4-7+3\sqrt{9-42\eta^4+81\eta^8}\right)^{1/3}.
\end{equation}
The value of $\alpha$ follows straightforwardly from the first of 
Eqs. (\ref{deltaalpha}).

\begin{figure}[thb]
\begin{center}
\includegraphics[clip,width=0.45\textwidth]{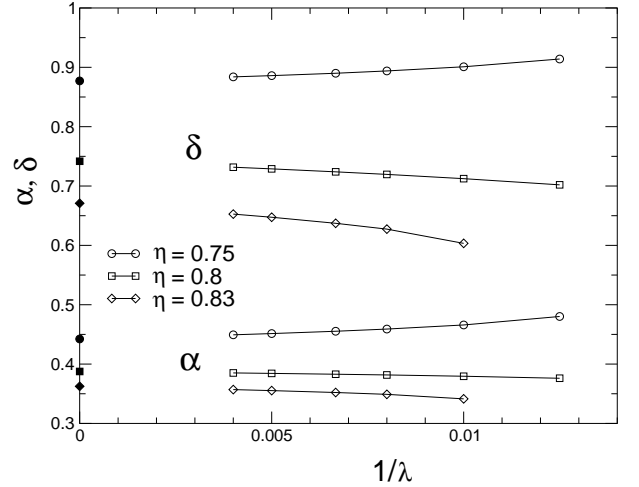}
\caption{The structure parameters $\delta$ and $\alpha$ of phase IVB 
vs. $\lambda$ for three values of $\eta=0.75,0.8,0.83$. 
Numerical data are represented by open symbols (connected by solid lines), 
the asymptotic $\lambda\to\infty$ result given by Eqs. (\ref{deltaalpha})
is depicted by full symbol.}
\label{fig:del-alf-lam}
\end{center}
\end{figure}

It is easy to check that the above formulas give the correct lattice
parameters at the endpoints of the phase IVB region, namely we have
$(\alpha=1/2,\delta=1)$ at $\eta=1/\sqrt{2}$ (phase III) and
$(\alpha=1/3,\delta=1/\sqrt{3})$ at $\eta=2/3^{3/4}$ (phase V). 
We did not find in the literature the above specification of 
the structural parameters of phase IVB in the hard-spheres limit.
The numerical test of the results for phase IV parameters
is depicted in Fig. \ref{fig:del-alf-lam}.
For a given $\eta$, the dependence of the parameters $\delta$ (top set)
and $\alpha$ (bottom set) on $1/\lambda$, obtained by numerics, is represented 
by open symbols (connected by solid line), the asymptotic $\lambda\to\infty$ 
result given by our Eqs. (\ref{deltaalpha}) is depicted by full symbol.
It is seen that numerical data converge quickly to their asymptotic values. 

\section{Conclusion} \label{Sec10}
In this paper, we have studied the zero-temperature phase diagram of 
bilayer Wigner crystals of Yukawa particles. 
To calculate the energy per particle of the phases, we used the recent method 
of lattice summations \cite{Samaj1} extended to Yukawa potentials.
The weak point of the method is that one has to know ahead the possible
phases from numerical simulations.
The strong point is that the truncation of the series of the generalized
Misra functions provides extremely precise estimates of the energy, 
e.g. the truncation at the 5th term provides the accuracy within 
17 decimal digits.

Another strong point of Misra functions is that they can be readily
expanded around the critical point, providing in this way closed-form
expressions for the critical lines between phases II-III (\ref{critline})
and III-IVA (\ref{crith}).
Only few Misra functions contribute in the equations for the critical
lines in the asymptotic Coulomb $\lambda\to 0$ and hard-spheres
$\lambda\to\infty$ limits. 
The characteristic feature of the Coulomb limit is the parabolic shape of 
the critical lines, see Eq. (\ref{le23}) with the corresponding plot
in Fig. \ref{fig:trans23ll} for the II-III phase transition and 
Eq. (\ref{le34a}) with the corresponding plot in Fig. \ref{fig:trans34all} 
for the III-IVA phase transition.
In the hard-spheres limit, the asymptotic formulas for the II-III
phase transition (\ref{e23as}) and the III-IVA phase transition (\ref{e34aas}) 
are pictures by dash-dotted lines in Fig. \ref{fig:phd2}.
It turns out that the second-order phase transitions II-III and 
III-IVA exhibit the mean-field critical exponents (\ref{MFind}). 

The most important features of the Yukawa phase diagram obtained 
by Messina and L\"owen \cite{Messina} were confirmed. 
On the contrary to previous suggestions, phase I goes directly to phase II
at $\eta=0$, i.e. there does not exist a finite interval of positive
$\eta$-values where phase I dominates.
Another important novelty is that instead of the suggested region of 
phase coexistence, we found a narrow channel within one continuous region 
of phase IVA.
This fact also lead to the tricritical point where 
the phases IVA, IVB and V meet.

Another application of our formalism is the determination of the structure
parameters of soft phases II and IVB in and close to the hard-spheres limit.
For $\lambda\to\infty$, the $\eta$-dependence of the aspect ratio $\Delta$ 
of phase II has already been known \cite{Messina}, see Eq. (\ref{DeltaHS}).
We were able to derive the first $1/\lambda$ correction to this 
asymptotic relation, see Eqs. (\ref{delta}) and (\ref{aeta}), which is 
in perfect agreement with the numerical results (Fig. \ref{fig:del-lam}).
The derivation of the $\eta$-dependence of two structure parameters 
$\delta$ and $\alpha$ of phase IVB in the limit $\lambda\to\infty$,
see the relations (\ref{exparg}) and Fig. \ref{fig:del-alf-lam},
is likely new as well.

As concerns future perspectives to apply our method to other systems, 
the system of particles with $1/r^{\sigma}$ interactions \cite{Mazars11} 
seems to be a good candidate. 

\begin{acknowledgments}
The support received from the grant VEGA No. 2/0015/2015 is acknowledged. 
\end{acknowledgments}

\appendix

\begin{widetext}

\section{}
We give explicit analytic formulas for several $z_{\nu}(x,y)$ functions
(\ref{znu}) with half-integer arguments:
\begin{eqnarray} 
z_{1/2}(x,y) & = & \sqrt{\frac{\pi}{x}}{\rm e}^{-2\sqrt{xy}}
\left[ 1-\frac{1}{2}\ {\rm erfc}{\left(\sqrt{\frac{x}{\pi}}-\sqrt{\pi y}\right)}
-\frac{1}{2}{\rm e}^{4\sqrt{xy}}\ 
{\rm erfc}{\left(\sqrt{\frac{x}{\pi}}+\sqrt{\pi y}\right)}\right], 
\label{z12} \\
z_{3/2}(x,y) & = & \sqrt{\frac{\pi}{y}}{\rm e}^{-2\sqrt{xy}}\left[ 1-\frac{1}{2}\ 
{\rm erfc}{\left(\sqrt{\frac{x}{\pi}}-\sqrt{\pi y}\right)}
+ \frac{1}{2}{\rm e}^{4\sqrt{xy}}\ 
{\rm erfc}{\left(\sqrt{\frac{x}{\pi}}+\sqrt{\pi y}\right)}\right], 
\label{z32} \\
z_{5/2}(x,y) & = & \frac{\sqrt{\pi x}}{y}{\rm e}^{-2\sqrt{xy}}
\left(1+\frac{1}{2\sqrt{xy}}\right)
-\frac{\sqrt{\pi}}{4y^{3/2}}\bigg[-4{\rm e}^{-x/\pi-\pi y}\sqrt{y} 
\nonumber \\ & & 
+{\rm e}^{-2\sqrt{xy}}\left(1+2\sqrt{xy}\right)\ 
{\rm erfc}{\left(\sqrt{\frac{x}{\pi}}-\sqrt{\pi y}\right)}
+{\rm e}^{2\sqrt{xy}}\left(-1+2\sqrt{xy}\right)\ 
{\rm erfc}{\left(\sqrt{\frac{x}{\pi}}+\sqrt{\pi y}\right)} \bigg], \label{z52ex}
\end{eqnarray}
\begin{eqnarray}
z_{7/2}(x,y) & = & \sqrt{\frac{\pi}{y^3}}\ x {\rm e}^{-2\sqrt{xy}}\left(1+
\frac{3}{2\sqrt{xy}}+\frac{3}{4xy}\right)
-\frac{\sqrt{\pi}}{8y^{5/2}}\left[-4{\rm e}^{-x/\pi-\pi y}(3+2\pi y)\sqrt{y}
\right. \nonumber\\ & & \left.
+{\rm e}^{-2\sqrt{xy}}\left(4xy+6\sqrt{xy}+3\right)\ 
{\rm erfc}{\left(\sqrt{\frac{x}{\pi}}-\sqrt{\pi y}\right)}
-{\rm e}^{2\sqrt{xy}}\left(4xy-6\sqrt{xy}+3\right)\ 
{\rm erfc}{\left(\sqrt{\frac{x}{\pi}}+\sqrt{\pi y}\right)} \right]. 
\label{z72ex}
\end{eqnarray}
\end{widetext}
Here, we introduced the complementary error function \cite{Gradshteyn}
\begin{equation} \label{erfc}
{\rm erfc}(z)=\frac{2}{\sqrt{\pi}}\int_z^\infty\exp{(-t^2)}\ {\rm d}t.
\end{equation}
The case $\nu=1/2$ can be found at the end of Ref. \cite{Chaudhry}.
The expressions for larger $\nu$ can be obtained by applying 
the obvious relation 
\begin{equation} \label{znuprol}
\frac{\partial z_\nu(x,y)}{\partial y} = -z_{\nu+1}(x,y).
\end{equation}

The Misra function case $z_{\nu}(0,y)$ \cite{Misra} should be understood 
in the sense of the limit $x\to 0$,
\begin{eqnarray}
\label{znu0y}
z_{1/2}(0,y) & = &\frac{2}{\sqrt{\pi}}\left[{\rm e}^{-\pi y} -\pi \sqrt{y} \ 
{\rm erfc}{\left(\sqrt{\pi y}\right)}\right], \nonumber\\
z_{3/2}(0,y) & = &\sqrt{\frac{\pi}{y}}\ {\rm erfc}{\left(\sqrt{\pi y}\right)},
\nonumber\\
z_{5/2}(0,y) & = &\frac{\sqrt{\pi}}{2 y^{3/2}}\left[2 {\rm e}^{-\pi y}\sqrt{y} \ 
+{\rm erfc}{\left(\sqrt{\pi y}\right)}\right], \nonumber\\
z_{7/2}(0,y) & = &\frac{\sqrt{\pi}}{4 y^{5/2}}
\bigg[2 {\rm e}^{-\pi y}\sqrt{y}\left(3+2\pi y\right) \nonumber\\
& & + 3\ {\rm erfc}{\left(\sqrt{\pi y}\right)}\bigg].
\end{eqnarray}
We need also the asymptotic expansions of $z_{\nu}(x,y)$ when one of 
the arguments $x$ or $y$ is large.
For $x$ finite and $y\gg 1$, we get
\begin{eqnarray}
z_{1/2}(x,y)&=& \frac{{\rm e}^{-\pi y-x/\pi}}{y \pi^{3/2}}\left[1+{\cal O}\left(
\frac{1}{y}\right)\right], \nonumber\\
z_{3/2}(x,y)&=& \frac{{\rm e}^{-\pi y-x/\pi}}{y \sqrt{\pi}}\left[1+{\cal O}\left(
\frac{1}{y}\right)\right], \nonumber\\
z_{5/2}(x,y)&=& \sqrt{\pi}\frac{{\rm e}^{-\pi y-x/\pi}}{y}\left[1+{\cal O}\left(
\frac{1}{y}\right)\right], \nonumber\\
z_{7/2}(x,y)&=& \sqrt{\pi^3}\frac{{\rm e}^{-\pi y-x/\pi}}{y}\left[1+{\cal O}\left(
\frac{1}{y}\right)\right]. \label{znuasym}
\end{eqnarray}
For $y$ finite and $x\gg 1$, we have
\begin{eqnarray}
z_{1/2}(x,y) & = & \sqrt{\frac{\pi}{x}} {\rm e}^{-2\sqrt{xy}}
+{\cal O}\left(\frac{1}{x}{\rm e}^{-x/\pi}\right), \nonumber\\
z_{3/2}(x,y) & = & \sqrt{\frac{\pi}{y}} {\rm e}^{-2\sqrt{xy}}
+{\cal O}\left(\frac{1}{x}{\rm e}^{-x/\pi}\right), \nonumber\\
z_{5/2}(x,y) & = & \frac{\sqrt{\pi x}}{y}{\rm e}^{-2\sqrt{xy}}\left(1+
\frac{1}{2\sqrt{xy}}\right)
+{\cal O}\left( \frac{1}{x} {\rm e}^{-x/\pi}\right), \nonumber\\
z_{7/2}(x,y) & = & \sqrt{\frac{\pi}{y^3}}\ x {\rm e}^{-2\sqrt{xy}}\left(1+
\frac{3}{2\sqrt{xy}}+\frac{3}{4xy}\right) \nonumber\\
& & +{\cal O}\left( \frac{1}{x} {\rm e}^{-x/\pi}\right). \label{znuasymp}
\end{eqnarray}
We applied the large-argument expansion of the error function 
\cite{Gradshteyn}, erfc$(z)\approx\exp({-z^2})/(\sqrt{\pi}z)$.

For small arguments $\delta x$ and $\delta y$, we shall need 
the following expansion
\begin{eqnarray}
z_\nu \left(x+\delta x,y+\delta y\right) = 
\int_0^{1/\pi} \frac{{\rm d}t}{t^\nu}{\rm e}^{-(x+\delta x)t}{\rm e}^{-(y+\delta y)/t}
\nonumber\\
\approx \int_0^{1/\pi} \frac{{\rm d}t}{t^\nu}{\rm e}^{-x t}{\rm e}^{-y/t}
(1-\delta x\ t)(1-\delta y/t) \nonumber\\
\approx z_\nu(x,y)-\delta x\ z_{\nu-1}(x,y) -\delta y\ z_{\nu+1}(x,y), 
\phantom{aa} \label{znusmall}
\end{eqnarray}
where we kept only terms linear in small variables.

\begin{widetext}

\section{}
The coefficients $f_{1,2}(\eta,\lambda)$ in Eq. (\ref{e21}) are given by
\begin{eqnarray}
f_1(\eta,\lambda) & = & \frac{\eta}{2\sqrt{\pi}\lambda}
\Bigg\{ 2\sum_{j=1}^\infty\left[
j^2 z_{5/2}\left(0,\frac{\lambda^2}{4 \pi^2 \eta^2}+j^2\sqrt{3}\right)
-\frac{j^2}{3} z_{5/2}\left(0,\frac{\lambda^2}{4 \pi^2 \eta^2}
+\frac{j^2}{\sqrt{3}}\right) \right] \nonumber\\ 
& & + 4\sum_{j,k=1}^\infty \left(k^2-\frac{j^2}{3}\right) 
z_{5/2}\left(0,\frac{\lambda^2}{4\pi^2\eta^2}+\frac{j^2}{\sqrt{3}}
+k^2\sqrt{3}\right) \nonumber\\ 
& & 2\sum_{j=1}^\infty (-1)^j\left[ j^2 
z_{5/2}\left(\pi^2\eta^2,\frac{\lambda^2}{4 \pi^2 \eta^2}+j^2\sqrt{3}\right)
-\frac{j^2}{3} z_{5/2}\left(\pi^2\eta^2,\frac{\lambda^2}{4 \pi^2 \eta^2}
+\frac{j^2}{\sqrt{3}} \right)\right] \nonumber\\ 
& & + 4\sum_{j,k=1}^\infty (-1)^j(-1)^k\left( k^2-\frac{j^2}{3}\right) 
z_{5/2}\left(\pi^2\eta^2,\frac{\lambda^2}{4 \pi^2 \eta^2}
+\frac{j^2}{\sqrt{3}}+k^2{\sqrt{3}}\right) \nonumber\\ 
& & + 2\sum_{j=1}^\infty\left[ j^2 
z_{5/2}\left(\frac{\lambda^2}{4 \eta^2},j^2 \sqrt{3}\right) -\frac{j^2}{3} 
z_{5/2}\left(\frac{\lambda^2}{4 \eta^2},\frac{j^2}{\sqrt{3}}\right)\right]
\nonumber\\ & &
+ 4\sum_{j,k=1}^\infty \left(k^2-\frac{j^2}{3}\right) 
z_{5/2}\left(\frac{\lambda^2}{4 \eta^2},\frac{j^2}{\sqrt{3}}+k^2{\sqrt{3}}\right)
\nonumber\\ & & 4\sum_{j,k=1}^\infty \left[(k-1/2)^2-\frac{(j-1/2)^2}{3}\right] 
z_{5/2}\left[\frac{\lambda^2}{4 \eta^2},\eta^2+
\frac{(j-1/2)^2}{\sqrt{3}}+(k-1/2)^2\sqrt{3}\right] \Bigg\}, \label{f1}
\end{eqnarray}
\begin{eqnarray}
f_2(\eta,\lambda) & = & \frac{\eta}{2\sqrt{\pi}\lambda}
\Bigg( 2\sum_{j=1}^\infty\left[ \frac{j^4}{18} 
z_{7/2}\left(0,\frac{\lambda^2}{4 \pi^2 \eta^2}+\frac{j^2}{\sqrt{3}}\right)
+\frac{j^4}{2} z_{7/2}\left(0,\frac{\lambda^2}{4 \pi^2 \eta^2}+j^2\sqrt{3}\right)
-\frac{j^2}{3\sqrt{3}} z_{5/2}\left(0,\frac{\lambda^2}{4 \pi^2 \eta^2}
+\frac{j^2}{\sqrt{3}}\right) \right] \nonumber \\ 
& & + 4\sum_{j,k=1}^\infty\left[\frac{1}{2}\left(k^2-\frac{j^2}{3}\right)^2 
z_{7/2}\left(0,\frac{\lambda^2}{4\pi^2\eta^2}+\frac{j^2}{\sqrt{3}}
+k^2\sqrt{3}\right)-\frac{j^2}{3\sqrt{3}} 
z_{5/2}\left(0,\frac{\lambda^2}{4 \pi^2 \eta^2}+\frac{j^2}{\sqrt{3}}
+k^2\sqrt{3}\right)\right] \nonumber \\ 
& & + 2\sum_{j=1}^\infty(-1)^j\bigg[ \frac{j^4}{18} 
z_{7/2}\left(\pi^2\eta^2,\frac{\lambda^2}{4 \pi^2 \eta^2}
+\frac{j^2}{\sqrt{3}}\right) + \frac{j^4}{2} 
z_{7/2}\left(\pi^2\eta^2,\frac{\lambda^2}{4 \pi^2 \eta^2}+j^2\sqrt{3}\right) 
\nonumber\\ & & - \frac{j^2}{3\sqrt{3}} 
z_{5/2}\left(\pi^2\eta^2,\frac{\lambda^2}{4 \pi^2\eta^2}
+\frac{j^2}{\sqrt{3}}\bigg) \right] \nonumber\\ 
& & + 4\sum_{j,k=1}^\infty (-1)^j(-1)^k\bigg[
\frac{1}{2}\left(k^2-\frac{j^2}{3}\right)^2
z_{7/2}\left(\pi^2\eta^2,\frac{\lambda^2}{4 \pi^2 \eta^2}
+\frac{j^2}{\sqrt{3}}+k^2{\sqrt{3}}\right) \nonumber\\ 
& & - \frac{j^2}{3\sqrt{3}}
z_{5/2}\left(\pi^2\eta^2,\frac{\lambda^2}{4 \pi^2 \eta^2}
+\frac{j^2}{\sqrt{3}}+k^2{\sqrt{3}}\right)\bigg] \nonumber \\ 
& & 2\sum_{j=1}^\infty\bigg[
\frac{j^4}{18} z_{7/2}\left(\frac{\lambda^2}{4 \eta^2},\frac{j^2}{\sqrt{3}}\right)
+\frac{j^4}{2} z_{7/2}\left(\frac{\lambda^2}{4 \eta^2},j^2\sqrt{3}\right)
-\frac{j^2}{3\sqrt{3}} 
z_{5/2}\left(\frac{\lambda^2}{4 \eta^2},\frac{j^2}{\sqrt{3}}\right) \bigg]
\nonumber\\ & & 
+ 4\sum_{j,k=1}^\infty \bigg[\frac{1}{2}\left(k^2-\frac{j^2}{3}\right)^2
z_{7/2}\left(\frac{\lambda^2}{4 \eta^2},\frac{j^2}{\sqrt{3}}+k^2{\sqrt{3}}\right)
- \frac{j^2}{3\sqrt{3}}
z_{5/2}\left(\frac{\lambda^2}{4 \eta^2},\frac{j^2}{\sqrt{3}}+k^2{\sqrt{3}}\right)
\bigg] \nonumber \\ & & 4\sum_{j,k=1}^\infty \bigg\{ \frac{1}{2}
\left[(k-1/2)^2-\frac{(j-1/2)^2}{3}\right]^2 
z_{7/2}\left[\frac{\lambda^2}{4 \eta^2},\eta^2+
\frac{(j-1/2)^2}{\sqrt{3}}+(k-1/2)^2\sqrt{3}\right] \nonumber \\ 
& & -\frac{(j-1/2)^2}{3\sqrt{3}} 
z_{5/2}\left[\frac{\lambda^2}{4 \eta^2},\eta^2+
\frac{(j-1/2)^2}{\sqrt{3}}+(k-1/2)^2\sqrt{3}\right]\bigg\} \Bigg). \label{f2}
\end{eqnarray}

We are interested in the small-$\eta$ behavior of the above functions.
One of the arguments in the $z_\nu(x,y)$ functions becomes large,
thus we can apply the asymptotic relations (\ref{znuasym}) and (\ref{znuasymp}).
Neglecting the exponentially small terms we find that only
the seventh and the ninth (last) sums both in (\ref{f1}) and (\ref{f2})
contribute, namely the leading terms with $j=1$ and $j=k=1$, respectively.
Consequently, for a fixed $\lambda>0$ and $\eta\to 0$ 
(i.e. $\lambda/\eta \gg 1$), we have
\begin{eqnarray}
f_1(\eta,\lambda)&\approx&\frac{\eta}{2\sqrt{\pi}\lambda}\left[-\frac{2}{3}
z_{5/2}\left(\frac{\lambda^2}{4 \eta^2},\frac{1}{\sqrt{3}}\right)
+\frac{2}{3}
z_{5/2}\left(\frac{\lambda^2}{4 \eta^2},\eta^2+\frac{1}{\sqrt{3}}\right)\right]
\approx -\frac{1}{2\sqrt{3}}{\rm e}^{-\frac{\lambda}{3^{1/4}\eta}}
+\frac{1}{6\left( \eta^2+\frac{1}{\sqrt{3}}\right)}
{\rm e}^{-\frac{\lambda}{\eta}\sqrt{\eta^2+\frac{1}{\sqrt{3}}}} , \nonumber\\ 
f_2(\eta,\lambda)&\approx&\frac{\eta}{2\sqrt{\pi}\lambda}\left[\frac{1}{9}
z_{7/2}\left(\frac{\lambda^2}{4 \eta^2},\frac{1}{\sqrt{3}}\right) + \frac{1}{18}
z_{7/2}\left(\frac{\lambda^2}{4 \eta^2},\eta^2+\frac{1}{\sqrt{3}}\right)\right]
\nonumber\\ 
&\approx& -\frac{\lambda}{8\ 3^{5/4}\eta}{\rm e}^{-\frac{\lambda}{3^{1/4}\eta}}
+\frac{\lambda}{144\eta(\eta^2+\frac{1}{\sqrt{3}})^{3/2}}
{\rm e}^{-\frac{\lambda}{\eta}\sqrt{\eta^2+\frac{1}{\sqrt{3}}}}, \label{f1a}
\end{eqnarray}
where we repeatedly neglected subleading terms.
Expanding also the second exponential
$\exp[-\lambda\sqrt{1+\sqrt{3}\eta^2}/(3^{1/4}\eta)]$ in $\eta$, 
we get (\ref{f1f2}).

\section{}
The coefficient $g_2(\eta,\lambda)$ in Eq. (\ref{e2gl}) takes the form
\begin{eqnarray}
g_2(\eta,\lambda) & = &\frac{\eta}{\sqrt{\pi}\lambda}\Bigg( 
\sum_{j=1}^\infty\left[ j^4 
z_{7/2}\left(0,\frac{\lambda^2}{4 \pi^2 \eta^2}+j^2\right)
-j^2 z_{5/2}\left(0,\frac{\lambda^2}{4 \pi^2 \eta^2}+j^2\right)\right]
\nonumber\\ & & + \sum_{j,k=1}^\infty\left[
(j^2-k^2)^2 z_{7/2}\left(0,\frac{\lambda^2}{4 \pi^2 \eta^2}+j^2+k^2\right)
-(j^2+k^2) z_{5/2}\left(0,\frac{\lambda^2}{4 \pi^2 \eta^2}+j^2+k^2\right)\right]
\nonumber\\ & & + \sum_{j=1}^\infty (-1)^j\left[
j^4 z_{7/2}\left(\pi^2\eta^2,\frac{\lambda^2}{4 \pi^2 \eta^2}+j^2\right)
-j^2 z_{5/2}\left(\pi^2\eta^2,\frac{\lambda^2}{4 \pi^2 \eta^2}+j^2\right)\right]
\nonumber\\ & & + \sum_{j,k=1}^\infty (-1)^j(-1)^k\left[ (j^2-k^2)^2 
z_{7/2}\left(\pi^2\eta^2,\frac{\lambda^2}{4 \pi^2 \eta^2}+j^2+k^2\right)-(j^2+k^2) 
z_{5/2}\left(\pi^2\eta^2,\frac{\lambda^2}{4 \pi^2 \eta^2}+j^2+k^2\right)\right]
\nonumber\\ & & + \sum_{j=1}^\infty\left[
j^4 z_{7/2}\left(\frac{\lambda^2}{4 \eta^2},j^2\right)
-j^2 z_{5/2}\left(\frac{\lambda^2}{4 \eta^2},j^2\right)\right] \nonumber\\ 
& & + \sum_{j,k=1}^\infty \left[
(j^2-k^2)^2 z_{7/2}\left(\frac{\lambda^2}{4 \eta^2},j^2+k^2\right)
-(j^2+k^2) z_{5/2}\left(\frac{\lambda^2}{4 \eta^2},j^2+k^2\right)\right]
\nonumber\\ & & + \sum_{j,k=1}^\infty \bigg\{
\left[(j-1/2)^2-(k-1/2)^2\right]^2 
z_{7/2}\left[\frac{\lambda^2}{4 \eta^2},\eta^2+(j-1/2)^2+(k-1/2)^2\right]
\nonumber\\ & &
- \left[(j-1/2)^2+(k-1/2)^2\right] z_{5/2}\left[\frac{\lambda^2}{4 \eta^2},\eta^2
+(j-1/2)^2+(k-1/2)^2\right]\bigg\}\Bigg). \label{g2}
\end{eqnarray}

Our next step is to analyze the small $\lambda$ behavior of the critical line 
between phases II-III which is given by $g_2(\eta,\lambda)=0$; here, we write 
$(\eta,\lambda)$ instead of $(\eta^c,\lambda^c)$ to simplify the notation. 
There are two small quantities: $\lambda^2$ and $\eta-\eta_0$, 
where we denote the Coulomb transition distance
$\eta^c(0)\equiv \eta_0 \approx 0.262760268246823\ldots$.
Applying formula (\ref{znusmall}), we demonstrate one specific example of 
the expansion of the generalized Misra functions in (\ref{g2}), up to terms 
linear in small variables $\lambda^2$ and $\eta-\eta_0$:
\begin{equation} \label{znuC}
z_{7/2}\left(\pi^2\eta^2,\frac{\lambda^2}{4\pi^2\eta^2}+j^2\right)
\approx z_{7/2}\left(\pi^2\eta_0^2,j^2\right) - \frac{\lambda^2}{4\pi^2\eta_0^2}
z_{9/2}\left(\pi^2\eta_0^2,j^2\right) - 
2\pi^2\eta_0(\eta-\eta_0)z_{5/2}\left(\pi^2\eta_0^2,j^2\right).
\end{equation}
Here, we used that
$\eta^2=[\eta_0+(\eta-\eta_0)]^2\approx\eta_0^2+2\eta_0(\eta-\eta_0)$.
The absolute terms, like the leading one on the r.h.s. of (\ref{znuC}), 
are canceled by the definition of the critical point at $\lambda=0$:
$g_2(\eta_0,0)=0$. 
Thus we are left with 
\begin{equation} \label{lametac23}
c_1^{(23)} \lambda^2 + c_2^{(23)}(\eta-\eta_0) = 0 ,
\end{equation}
where
\begin{eqnarray}
c_1^{(23)} & = & - \sum_{j=1}^\infty\left[j^4 z_{9/2}\left(0,j^2\right)
-j^2 z_{7/2}\left(0,j^2\right)\right]/(4\pi^2\eta_0^2) \nonumber\\ 
& & -\sum_{j,k=1}^\infty\left[(j^2-k^2)^2 z_{9/2}\left(0,j^2+k^2\right)-
(j^2+k^2) z_{7/2}\left(0,j^2+k^2\right)\right]/(4\pi^2\eta_0^2) \nonumber\\ 
& & - \sum_{j=1}^\infty (-1)^j\left[j^4 z_{9/2}\left(\pi^2\eta_0^2,j^2\right)
-j^2 z_{7/2}\left(\pi^2\eta_0^2,j^2\right)\right]/(4\pi^2\eta_0^2) \nonumber\\ 
& & - \sum_{j,k=1}^\infty (-1)^j(-1)^k\left[(j^2-k^2)^2 
z_{9/2}\left(\pi^2\eta_0^2,j^2+k^2\right)
-(j^2+k^2) z_{7/2}\left(\pi^2\eta_0^2,j^2+k^2\right)\right]/(4\pi^2\eta_0^2)
\nonumber\\ & & - \sum_{j=1}^\infty\left[j^4 z_{5/2}\left(0,j^2\right)
-j^2 z_{3/2}\left(0,j^2\right)\right]/(4\eta_0^2) \nonumber\\ 
& & - \sum_{j,k=1}^\infty \left[ (j^2-k^2)^2 z_{5/2}\left(0,j^2+k^2\right)
-(j^2+k^2) z_{3/2}\left(0,j^2+k^2\right)\right]/(4\eta_0^2) \nonumber\\ 
& & - \sum_{j,k=1}^\infty \bigg\{ \left[(j-1/2)^2-(k-1/2)^2\right]^2 
z_{5/2}\left[0,\eta_0^2+(j-1/2)^2+(k-1/2)^2\right] \nonumber\\
& & + \left[(j-1/2)^2+(k-1/2)^2\right] 
z_{3/2}\left[0,\eta_0^2+(j-1/2)^2+(k-1/2)^2\right]\bigg\}/(4\eta_0^2)
\approx-0.04791591901052 \label{c123}
\end{eqnarray}
and
\begin{eqnarray}
c_2^{(23)} & = & - \sum_{j=1}^\infty (-1)^j\left[j^4 
z_{5/2}\left(\pi^2\eta_0^2,j^2\right)
-j^2 z_{3/2}\left(\pi^2\eta_0^2,j^2\right)\right]2\pi^2\eta_0 \nonumber \\ 
& & - \sum_{j,k=1}^\infty (-1)^j(-1)^k
\left[(j^2-k^2)^2 z_{5/2}\left(\pi^2\eta_0^2,j^2+k^2\right)
-(j^2+k^2) z_{3/2}\left(\pi^2\eta_0^2,j^2+k^2\right)\right]2\pi^2\eta_0
\nonumber \\ & & - \sum_{j,k=1}^\infty \bigg\{
\left[(j-1/2)^2-(k-1/2)^2\right]^2 
z_{9/2}\left[0,\eta_0^2+(j-1/2)^2+(k-1/2)^2\right] \nonumber\\
& & + \left[(j-1/2)^2+(k-1/2)^2\right] 
z_{7/2}\left[0,\eta_0^2+(j-1/2)^2+(k-1/2)^2\right]\bigg\}2\eta_0
\approx 1.15830861669576 . \label{c223}
\end{eqnarray}
Eq. (\ref{lametac23}) with the specified constants yields (\ref{le23}).

\section{}
The function $h_2(\eta,\lambda)$ in Eq. (\ref{e4agl}) reads as follows
\begin{eqnarray}
h_2(\eta,\lambda) & = & \frac{\eta}{2\sqrt{2\pi}\lambda}\Bigg( 
\sum_{j=1}^{\infty} \left[1 + (-1)^j\right]
\left[ j^4 z_{7/2}\left(0,\frac{\lambda^2}{2 \pi^2 \eta^2}+j^2\right)
-j^2 z_{5/2}\left(0,\frac{\lambda^2}{2 \pi^2 \eta^2}+j^2\right)\right]
\nonumber\\ & & + \sum_{j,k=1}^\infty\left[1 + (-1)^j(-1)^k\right]\left[
(j^2-k^2)^2 z_{7/2}\left(0,\frac{\lambda^2}{2 \pi^2 \eta^2}+j^2+k^2\right)
-(j^2+k^2) z_{5/2}\left(0,\frac{\lambda^2}{2 \pi^2 \eta^2}+j^2+k^2\right)\right]
\nonumber\\ & & + \sum_{j=1}^\infty \left[1 + (-1)^j\right]\left[
j^4 z_{7/2}\left(\eta^2\pi^2/2,\frac{\lambda^2}{2 \pi^2 \eta^2}+j^2\right)
-j^2 z_{5/2}\left(\eta^2\pi^2/2,\frac{\lambda^2}{2 \pi^2 \eta^2}+j^2\right)\right]
\nonumber\\ & & + 2 \sum_{j,k=1}^{\infty} (-1)^j \left[ (j^2-k^2)^2 
z_{7/2}\left(\frac{\eta^2\pi^2}{2},\frac{\lambda^2}{2 \pi^2 \eta^2}+j^2+k^2\right)
-(j^2+k^2) z_{5/2}\left(\frac{\eta^2\pi^2}{2},\frac{\lambda^2}{2 \pi^2 \eta^2}
+j^2+k^2\right)\right] \nonumber \\ & & + \sum_{j=1}^\infty \left[
j^4 z_{7/2}\left(\frac{\lambda^2}{2 \eta^2},j^2\right)
-j^2 z_{5/2}\left(\frac{\lambda^2}{2 \eta^2},j^2\right)\right]
\nonumber \\ & & + \sum_{j,k=1}^\infty\left[
(j^2-k^2)^2 z_{7/2}\left(\frac{\lambda^2}{2 \eta^2},j^2+k^2\right)
-(j^2+k^2) z_{5/2}\left(\frac{\lambda^2}{2 \eta^2},j^2+k^2\right)\right]
\nonumber \\ & & + \sum_{j,k=1}^\infty\bigg\{ \left[(j-1/2)^2-(k-1/2)^2\right]^2 
z_{7/2}\left[\frac{\lambda^2}{2 \eta^2},(j-1/2)^2+(k-1/2)^2\right] \nonumber \\ 
& & - \left[(j-1/2)^2+(k-1/2)^2\right] z_{5/2}\left[
\frac{\lambda^2}{2 \eta^2},(j-1/2)^2+(k-1/2)^2\right]\bigg\} \nonumber \\ 
& & + 2\sum_{j,k=1}^\infty \bigg\{
\left[(j-1/2)^2-k^2\right]^2 z_{7/2}\left[\frac{\lambda^2}{2 \eta^2},
\frac{\eta^2}{2}+(j-1/2)^2+k^2\right] \nonumber \\
& & - \left[(j-1/2)^2+k^2\right] z_{5/2}\left[\frac{\lambda^2}{2 \eta^2},
\frac{\eta^2}{2}+(j-1/2)^2+k^2\right]\bigg\} \nonumber\\ 
& & + \sum_{j=1}^\infty\left\{ (j-1/2)^4 z_{7/2}\left[\frac{\lambda^2}{2 \eta^2},
\frac{\eta^2}{2}+(j-1/2)^2\right] -(j-1/2)^2 
z_{5/2}\left[\frac{\lambda^2}{2 \eta^2},
\frac{\eta^2}{2}+(j-1/2)^2\right]\right\} \Bigg) . \label{h2}
\end{eqnarray}

Now we can analyze the low-$\lambda$ limit of the critical line 
between phases III and IVA. 
Proceeding in the same way as in the previous case of the II-III transition,
taking the value $\eta_0=\eta^c(0)\approx 0.621480924579783$,
we get from (\ref{h2}) the equality
\begin{equation} \label{lametac34}
c_1^{(34)} \lambda^2 + c_2^{(34)}(\eta-\eta_0) =  0
\end{equation}
with $c_1^{(34)}\approx 0.0063328359292865$
and $c_2^{(34)}\approx -0.94855575801235884369$,
so that Eq. (\ref{le34a}) follows.
\end{widetext}

\end{document}